\documentclass[lettersize,journal]{IEEEtran}

\usepackage{cite}
\usepackage{amsmath,amssymb,amsfonts}
\usepackage{graphicx}
\usepackage{textcomp}
\usepackage{xcolor}
\usepackage{authblk}

\usepackage{booktabs}
\usepackage{epsfig}
\usepackage{latexsym}
\usepackage{multirow}
\usepackage{stfloats}
\usepackage{epstopdf}
\usepackage{color}  
\usepackage{tabularx} 
\usepackage{enumerate}
\usepackage{array}
\graphicspath{{./Figures/}}
\usepackage{color}
\usepackage{bbm}
\usepackage{caption}
\usepackage{bm}
\usepackage[tight,footnotesize]{subfigure}
\usepackage{balance}
\usepackage{mathrsfs}
\usepackage{verbatim}
\allowdisplaybreaks[4]
\usepackage{dsfont}
\usepackage{verbatim}
\usepackage{tikz}
\usepackage{setspace}
\usepackage{diagbox}
\usepackage[framemethod=tikz]{mdframed}
\usepackage{multicol}
\usepackage{environ}
\usepackage{tikz}
\usepackage{stfloats}
\usepackage{algpseudocode}
\usepackage{graphics}
\usepackage{epsfig}
\usepackage{amsthm}
\usepackage{authblk}
\usepackage{enumitem}
\usepackage{makecell}
\newcommand{\rev}[1]{\textcolor{blue}{{#1}}}

\ifCLASSOPTIONcompsoc
\usepackage[caption=false, font=normalsize, labelfont=sf, textfont=sf]{subfig}
\else
\usepackage[caption=false, font=footnotesize]{subfig}

\usepackage{bm}
\def\BibTeX{{\rm B\kern-.05em{\sc i\kern-.025em b}\kern-.08em
    T\kern-.1667em\lower.7ex\hbox{E}\kern-.125emX}}
\begin{document}

\title{Cross Far- and Near-Field Channel Measurement and Modeling in Extremely Large-scale Antenna Array (ELAA) Systems}

\author{Yiqin~Wang,
Chong~Han,~\IEEEmembership{Senior~Member,~IEEE,}
Shu~Sun,
and~Jianhua~Zhang%
\thanks{Yiqin Wang is with Terahertz Wireless Communications (TWC) Laboratory, Shanghai Jiao Tong University, China (Email: wangyiqin@sjtu.edu.cn).}%
\thanks{Chong Han is with Terahertz Wireless Communications (TWC) Laboratory, also with Department of Electronic Engineering and the Cooperative Medianet Innovation Center (CMIC), Shanghai Jiao Tong University, China (Email: chong.han@sjtu.edu.cn).}%
\thanks{Shu Sun is with Department of Electronic Engineering and the Cooperative Medianet Innovation Center (CMIC), Shanghai Jiao Tong University, China (Email: shusun@sjtu.edu.cn).}%
\thanks{Jianhua Zhang is with the State Key Laboratory of Networking and Switching Technology, Beijing University of Posts and Telecommunications, Beijing, China (Email: jhzhang@bupt.edu.cn).}}

\maketitle
\begin{abstract}

Technologies like ultra-massive multiple-input-multiple-output (UM-MIMO) and reconfigurable intelligent surfaces (RISs) are of special interest to meet the key performance indicators of future wireless systems including ubiquitous connectivity and lightning-fast data rates. One of their common features, the extremely large-scale antenna array (ELAA) systems with hundreds or thousands of antennas, give rise to near-field (NF) propagation and bring new challenges to channel modeling and characterization.
In this paper, a cross-field channel model for ELAA systems is proposed, which improves the statistical model in 3GPP TR~38.901 by refining the propagation path with its first and last bounces and differentiating the characterization of parameters like path loss, delay, and angles in near- and far-fields. A comprehensive analysis of cross-field boundaries and closed-form expressions of corresponding NF or FF parameters are provided. Furthermore, cross-field experiments carried out in a typical indoor scenario at 300~GHz verify the variation of MPC parameters across the antenna array, and demonstrate the distinction of channels between different antenna elements. Finally, detailed generation procedures of the cross-field channel model are provided, based on which simulations and analysis on NF probabilities and channel coefficients are conducted for $4\times4$, $8\times8$, $16\times16$, and $9\times21$ uniform planar arrays at different frequency bands.

\end{abstract}

\section{Introduction}

\par In light of enormous spatial multiplexing and beamforming gain, ultra-massive multiple-input-multiple-output (UM-MIMO) and reconfigurable intelligent surfaces (RISs) are of special interests since they effectively increase the communication range and enhance the capacity to meet the key performance indicators of sixth generation (6G) communications~\cite{akyildiz2016realizing,saad2020vision}.
These technologies both feature extremely large-scale antenna array (ELAA) systems that are equipped with hundreds or thousands of antennas. For such a large aperture size, different from massive MIMO for fifth generation (5G) communications, new features like near-field (NF) propagation and spatial non-stationarity arise and bring new challenges to channel modeling and characterization~\cite{cui2022near,channel_tutorial}.

\par In wireless communications, the electromagnetic (EM) radiation field is divided into the far-field (FF) region and the radiative NF region\footnote{The near-field region is subdivided into reactive NF and radiative NF. Since EM waves in the reactive NF are induced fields and do not set off from the antenna, we only refer to the radiative NF in this article.}.
The basis of the near- and far-field (NF-FF) demarcation is various. In the FF region, the EM field is usually approximated by planar waves, whereas the planar wavefront model (PWM) may fail for wireless communication systems with ELAAs. Therefore, most boundaries are defined by the impact of PWM on channel characteristics such as amplitude and phase. For instance, the most classic NF-FF boundary, Rayleigh distance~\cite{selvan2017fraunhofer}, is defined by the distance between transmitter (Tx) and receiver (Rx) beyond which the maximum phase error across the antenna aperture by using the PWM is less than $\pi/8$.
Furthermore, by considering different arrival angles across elements in the antenna array, the variation of the projected aperture is taken into account in NF-FF demarcation~\cite{lu2022communicating}.
Besides, other criteria for NF-FF demarcation are based on the impact on beamforming~\cite{bjornson2021primer,cui2021near} or system performance such as channel capacity~\cite{lu2022communicating,sun2023how}.

\par Despite these researches on NF-FF demarcation, the effect of cross near- and far-fields has not been implemented in MIMO channel models with ELAA systems.
In this paper, we adapt the cluster-based statistical channel model in 3GPP TR~38.901~\cite{3gpp38901} to cross-field characterization of MIMO channels.
Specifically, as scatterers are likely to be located in near- or far-field of the antenna array at Tx or Rx, we divide EM wave propagation into three parts, separated by its first and last bounce. For the propagation from the Tx to the first bounce scatterer (FBS), and that from the last bounce scatterer (LBS) to the Rx, multi-path component (MPC) parameters including propagation time and departure/arrival angles are characterized in near- and far-fields, respectively.

\par Since far-field and near-field MPC parameters are generated separately, the NF-FF demarcation is first implemented by comparing the NF-FF boundary with the relative location of the scatterer to the antenna array. 
Specially, the NF-FF boundary is not shared by all MPC parameters. Instead, each parameter defines its own NF-FF boundary, on account of the effect of its approximation by PWM on channel coefficients.


\par The main contributions of this work are summarized as follows:
\begin{itemize}
    \item A cross far- and near-field channel model for ELAA systems is developed. Based on the refined model framework where the EM wave propagation is divided by FBS and LBS, we improve the cluster-based statistical channel model in 3GPP TR~38.901 by differentiating the characterization of parameters including path loss, propagation time and departure/arrival angles in FF and NF.
    \item Confirmatory cross-field experiments are carried out in a typical indoor scenario at 300~GHz. Equipped with a displacement platform and a rotator, the sounding system supports angular-resolved measurement of a virtual planar array with sub-millimeter precision. The analysis of the measurement result verifies the variation of MPC parameters across the antenna array, and demonstrates the distinction of channels between different antenna elements in the scenario where scatterers are located in the cross-field of the array.
    \item Detailed generation procedure of the cross-field channel model for ELAA systems is provided, based on the comprehensive analysis of cross-field boundaries and closed-form characterization of parameters. Channel simulations and analysis on NF probabilities and channel coefficients for $4\times4$, $8\times8$, $16\times16$, and $9\times21$ uniform planar arrays (UPAs) at sub-6~GHz and mmWave frequency bands are then carried out in a UMa scenario.
\end{itemize}
\par The remainder of this paper is organized as follows.
The cross-field MIMO channel model framework is first introduced in Section~\ref{sec:framework}, followed by the analysis of closed-form cross-field boundaries and characterization of parameters in Section~\ref{sec:analysis}. In Section~\ref{sec:campaign}, a probative measurement with virtual uniform planar array at 300~GHz is carried out and the measurement results are discussed. We then elaborate the implementation of the cross-field MIMO channel model and the simulation result in Section~\ref{sec:simulation}. Finally, the paper is concluded in Section~\ref{sec:conclusion}.

\section{Cross-field Channel Model Framework} \label{sec:framework}


Before jumping into the analysis of NF-FF boundaries for each parameter, the channel model framework based on twin scatterers is first introduced. This starts with the notations in the single-input-single-output (SISO) channel model, which is then extended to the proposed MIMO channel model.

\subsection{Channel Model based on Twin Scatterers}

\par A SISO channel model, simplified by omitting the change of channel coefficients caused by time variation and frequency offset of the subcarrier, is expressed as
\begin{equation}
    h(\tau,\theta) = \sum_{n=1}^{N}\sum_{m=1}^{M_n} g_{n,m} \cdot e^{-j2\pi ft} \cdot \delta(\tau-\tau_{n,m})\cdot\delta(\theta-\theta_{n,m}),
\end{equation}
where there are one line-of-sight (LoS) cluster ($n=1$) and $N-1$ non-line-of-sight (NLoS) clusters ($n=1,2,...,N$) from the transmitter (Tx) to the receiver (Rx).
\par Each NLoS cluster is related with a pair of scatterers, named by \textit{twin scatterers}, i.e., one first-bounce scatterer (FBS) and one last-bounce scatterer (LBS). As a result, each NLoS path is divided into three parts, represented by the vector from Tx to FBS $\vec{r}_{n}^{\rm(T)}$, the vector from FBS to LBS $\vec{r}_{n}^{\rm(V)}$, and the vector from LBS to Rx $\vec{r}_{n}^{\rm(R)}$. In the $n^\mathrm{th}$ cluster, there are $M_n$ paths (e.g., for the LoS path, $K_1=1$). The delay of each path is denoted as $\tau_{n,m}$, while $\Theta_{n,m}=\left[\varphi_{n,m}^{\rm(T)},\theta_{n,m}^{\rm(T)},\varphi_{n,m}^{\rm(R)},\theta_{n,m}^{\rm(R)}\right]$ indicates the azimuth angle of departure (AoD), zenith angle of departure (ZoD), azimuth angle of arrival (AoA) and zenith angle of arrival (ZoA) of the path. $g_{n,m}$ represents the path gain.
\par We first consider the cluster parameter. Due to the division of the propagation path, the delay of the cluster is also composed of three parts correspondingly, i.e., $\tau_{n}=\tau_{n}^{\rm(T)}+\tau_{n}^{\rm(V)}+\tau_{n}^{\rm(R)}$.
The first part from Tx to FBS determines $\tau_{n}^{\rm(T)}$, as well as angles of departure $\varphi_{n,m}^{\rm(T)}$ and $\theta_{n,m}^{\rm(T)}$. $\tau_{n}^{\rm (T)}=r_{n}^{\rm(T)}/c$ where $r_{n}^{\rm(T)}=\|\vec{r}_{n}^{\rm(T)}\|$ and $c$ denotes the speed of light. Take the position of Tx as the origin, the spherical coordinate ($r_{n}^{\rm(T)}$, $\theta_{n}^{\rm(T)}$, $\varphi_{n}^{\rm(T)}$) defines the position of FBS relative to Tx, which is equivalent to $\vec{r}_{n}^{\rm(T)}$.
For the second part, the virtual path between FBS and LBS, we only consider the influence of distance $r_{n}^{\rm(V)}=\|\vec{r}_{n}^{\rm(V)}\|$ on the second term in the cluster delay by $\tau_{n}^{\rm(V)}=r_{n}^{\rm(V)}/c$, while the direction is neglected.
The third part from LBS to Rx determines $\tau_{n}^{\rm(R)}$, as well as angles of arrival $\varphi_{n,m}^{\rm(R)}$ and $\theta{n,m}^{\rm(R)}$. $\tau_{n}^{\rm(R)}=r_{n}^{\rm(R)}/c$ where $r_{n}^{\rm(R)}=\|\vec{r}_{n}^{\rm(R)}\|$. The spherical coordinate ($r_{n}^{\rm(R)}$, $\theta_{n}^{\rm(R)}$, $\varphi_{n}^{\rm(R)}$) defines the position of LBS relative to Rx, which is equivalent to $\vec{r}_{n}^{\rm(R)}$.
The cluster gain is denoted as $g_{n}=\sqrt{P_n}$.
\par For each path in the cluster, the cluster power is equally divided, i,e, the path gain is $g_{n,m}=g_{n}/\sqrt{K}$ for $k=1,2,...,K_{n}$. The delay and angles of each path are generated based on the corresponding cluster parameter and intra-cluster spreads, by following the procedures in 3GPP TR 38.901.


\subsection{Multiple-Input-Multiple-Output Channel Model}
\begin{figure}
    \centering
    \includegraphics[width=\linewidth]{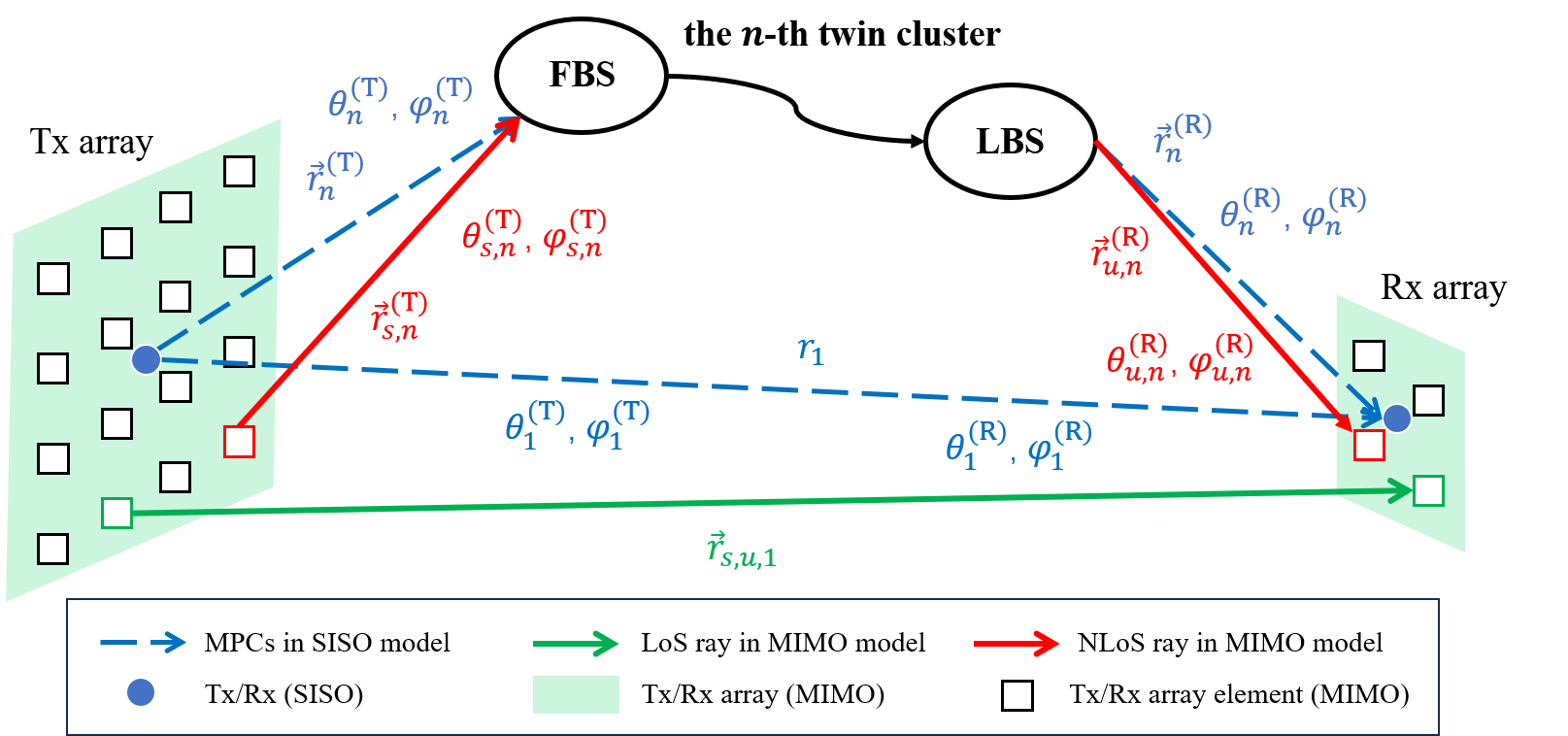}
    \caption{Twin-scatterer-based SISO and MIMO channels.}
    \label{fig:mimo}
\end{figure}

\par Before path generation inside each cluster, we need to consider different cluster parameters at each element. Therefore, we introduce the index for Tx element, $s$, and the index for Rx element, $u$. Then, we need to extend the clusters' delay $\tau_n$ departure angle ($\theta_{n}^{\rm(T)}$, $\varphi_{n}^{\rm(T)}$), arrival angle ($\theta_{n}^{\rm(R)}$, $\varphi_{n}^{\rm(R)}$), gain $g_n$, and number of path $K_n$ in the SISO model to cluster parameters at different antenna elements, $\tau_{s,u,n}$, $\theta_{s,u,n}^{\rm(T)}$, $\varphi_{s,u,n}^{\rm(T)}$, $\theta_{s,u,n}^{\rm(R)}$, $\varphi_{s,u,n}^{\rm(R)}$, $g_{s,u,n}$ and $K_{s,u,n}$.

\begin{table}
    \centering
    \caption{Symbols in the cross-field MIMO channel model.}
    \begin{tabular}{|c|c|}
         \hline
         \textbf{Symbol} & \textbf{Meaning} \\ \hline
         $n$ & Index of cluster \\ \hline
         $m$ & Index of path \\ \hline
         $s$ & Index of antenna element at Tx \\ \hline
         $u$ & Index of antenna element at Rx \\ \hline
         $f$ & Carrier frequency \\ \hline
         $\lambda$ & Wavelength \\ \hline
         $K_R$ & Ricean K-factor \\ \hline
         $\tau$ & Time of arrival (delay) \\ \hline
         $P$ & Power \\ \hline
         $\Omega = \left[\theta, \varphi\right]$ & Zenith and azimuth angles \\ \hline
         $F$ & Antenna pattern \\ \hline
         $\kappa$ & Cross-polarization power ratio \\ \hline
         $\hat{\mathbbm{r}}(\cdot)$ & Unit vector at the direction \\ \hline
         $\mathbbm{v}_{{\rm rx}}$ & Motion of Rx\\ \hline
    \end{tabular}
    \label{tab:model_symbol}
\end{table}

Refer to the channel model for large antenna array in Section~7.6.2 of 3GPP TR~38.901, the general form of the cross-field MIMO channel model is composed of channel impulse responses of element-to-element links, as
\begin{equation}
\begin{aligned}
    h_{u,s}(t,\tau;f) = \sqrt{\frac{K_R}{1+K_R}} \cdot &h_{u,s,{\rm LoS}}(t,f) \cdot \delta(\tau-\tau_{u,s,{\rm LoS}}) + \\
    \sqrt{\frac{1}{1+K_R}} \cdot \sum_{n=1}^{N}\sum_{m=1}^{M_{u,s,n}} &h_{u,s,n,m}(t,f) \cdot \delta(\tau-\tau_{u,s,n,m})
\end{aligned}
\end{equation}
where
\begin{equation}
\begin{aligned}
    h_{u,s,\rm LoS}(t,f) =& \sqrt{P_{1,{\rm LoS}}} \begin{bmatrix}F_{{\rm rx}}^{\hat{\theta}}(\Omega_{u,{\rm LoS}}^{\rm(R)})\\F_{{\rm rx}}^{\hat{\varphi}}(\Omega_{u,{\rm LoS}}^{\rm(R)})\end{bmatrix}^{\rm T} \\
    &\begin{bmatrix}1&0\\0&-1\end{bmatrix} \begin{bmatrix}F_{{\rm tx}}^{\hat{\theta}}(\Omega_{s,{\rm LoS}}^{\rm(T)})\\F_{{\rm tx}}^{\hat{\varphi}}(\Omega_{s,{\rm LoS}}^{\rm(T)})\end{bmatrix} \\
    &\exp\left(j\frac{2\pi t}{\lambda(f)} \mathbbm{v}_{{\rm rx}}^{\rm T} \hat{\mathbbm{r}}(\Omega_{u,{\rm LoS}}^{\rm(R)})\right)
\end{aligned}
\end{equation}
and
\begin{equation}
\begin{aligned}
    h_{u,s,n,m}(t,f) =&\begin{bmatrix}F_{{\rm rx}}^{\hat{\theta}}(\Omega_{u,n,m}^{\rm(R)})\\F_{{\rm rx}}^{\hat{\varphi}}(\Omega_{u,n,m}^{\rm(R)})\end{bmatrix}^{\rm T} \\
    &\begin{bmatrix}\exp(j\psi_{n,m}^{\hat{\theta}\hat{\theta}})&\kappa_{n,m}^{-1/2}\exp(j\psi_{n,m}^{\hat{\theta}\hat{\varphi}})\\\kappa_{n,m}^{-1/2}\exp(j\psi_{n,m}^{\hat{\varphi}\hat{\theta}})&\exp(j\psi_{n,m}^{\hat{\varphi}\hat{\varphi}})\end{bmatrix} \\
    &\begin{bmatrix}F_{{\rm tx}}^{\hat{\theta}}(\Omega_{s,n,m}^{\rm(T)})\\F_{{\rm tx}}^{\hat{\varphi}}(\Omega_{s,n,m}^{\rm(T)})\end{bmatrix} \\
    &\sqrt{\frac{P_n}{M_{u,s,n}}} \cdot \exp\left(j\frac{2\pi t}{\lambda(f)} \mathbbm{v}_{{\rm rx}}^{\rm T} \hat{\mathbbm{r}}(\Omega_{u,n,m}^{\rm(R)})\right)
\end{aligned}
\end{equation}

\par We apply the twin-scatterer-based framework and separate the propagation path into three parts. The first part goes from the $s^\mathrm{th}$ element at Tx to the $n$$^\mathrm{th}$ FBS, which is characterized by distance $r_{s,n}^{\rm(T)}$, azimuth angle of departure $\varphi_{s,n}^{\rm(T)}$ and zenith angle of departure $\theta_{s,n}^{\rm(T)}$.
The last part goes from the $n$$^\mathrm{th}$ LBS to the $u^\mathrm{th}$ element at Rx, which is characterized by distance $r_{u,n}^{\rm(R)}$, azimuth angle of departure $\varphi_{u,n}^{\rm(R)}$ and zenith angle of departure $\theta_{u,n}^{\rm(R)}$.
In the middle of the $n^\mathrm{th}$ twin scatterer, the link from FBS to LBS is regarded as a virtual path, whose path length is fixed by $r_{n}^{\rm(V)}$. Therefore, the total time of arrival from the ($m_{\rm T}$, $n_{\rm T}$)$^\mathrm{th}$ element at Tx to the $u^\mathrm{th}$ element at Rx is $\tau_{s,u,n}=r_{s,u,n}/c=(r_{s,n}^{\rm(T)}+r_{n}^{\rm(V)}+r_{u,n}^{\rm(R)})/c$, and the departure and arrival angles are indicated by $\Theta_{s,u,n} = \left[ \varphi_{s,n}^{\rm(T)}, \theta_{s,n}^{\rm(T)}, \varphi_{u,n}^{\rm(R)}, \theta_{u,n}^{\rm(R)} \right]$.

\subsection{Cross-Field MIMO Channel Modeling}
\par The major difference between NF and FF MIMO channel is the characterization of path parameters, which is array-level under FF assumption and element-level in the NF case.
To model the parameters in the cross-field, the evaluation of parameters is different based on whether a scatterer is in the NF or FF region of an antenna array. The partition is based on the comparison of the NF/FF boundary and the distance between the scatter and the array center.
\par First, since the propagation path is divided into three parts by Tx, FBS, LBS and Rx, NF/FF regions are distinguished respectively for parameters between the FBS and Tx array (for instance $\tau{s,n}^{\rm(T)}$, $\varphi_{s,n}^{\rm(T)}$, $\theta_{s,n}^{\rm(T)}$), and those between the LBS and the Rx array (for instance $\tau{s,n}^{\rm(R)}$, $\varphi_{u,n}^{\rm(R)}$, $\theta_{u,n}^{\rm(R)}$).
\par Second, different parameters, i.e., path length and angles, do not share the same NF/FF boundary. Instead, the NF-FF boundary is specific for each parameter and its corresponding approximation method.
In particular, for each approximation method of a parameter, we first calculate the magnitude/phase error between the approximation and the ground truth given by the spherical-wave model (SWM). Then, we derive the corresponding NF-FF boundary by limiting this error to the maximum tolerance of magnitude/phase error.
\section{Large- and Small-scale Parameters in Cross-Field} \label{sec:analysis}

In this section, we determine the large- and small-scale parameters in the cross-field and derive respective NF-FF boundaries for each parameter.
The following analysis is by default specific to parameters between FBS and a $M_{\rm T} \times N_{\rm T}$ UPA at Tx, with element spacing $d_{\rm T}$. The index of antenna element $s$ can be equivalently replaced by ($m_{\rm T}$,$n_{\rm T}$). The analysis for other antenna array types, like uniform linear array (ULA) and uniform circular array (UCA) is also provided in this section.
The analysis is similar for parameters between LBS and arrays at Rx.

\subsection{Delay in Cross-Field}

\par The following discussion mainly focuses on the approximation of the propagation length from the ($m_{\rm T}$, $n_{\rm T}$)$^\mathrm{th}$ element ($m_{\rm T}=1,2,...,M_{\rm T}$, $n_{\rm T}=1,2,...,N_{\rm T}$) in Tx-UPA to the $n^\mathrm{th}$ FBS, $r_{m_{\rm T},n_{\rm T},n}^{\rm(T)}$. The corersponding delay (or propagation time) can be derived by $\tau_{m_{\rm T},n_{\rm T},n}^{\rm(T)}=r_{m_{\rm T},n_{\rm T},n}^{\rm(T)}/c$.

\begin{figure}
    \centering
    \includegraphics[width=0.8\linewidth]{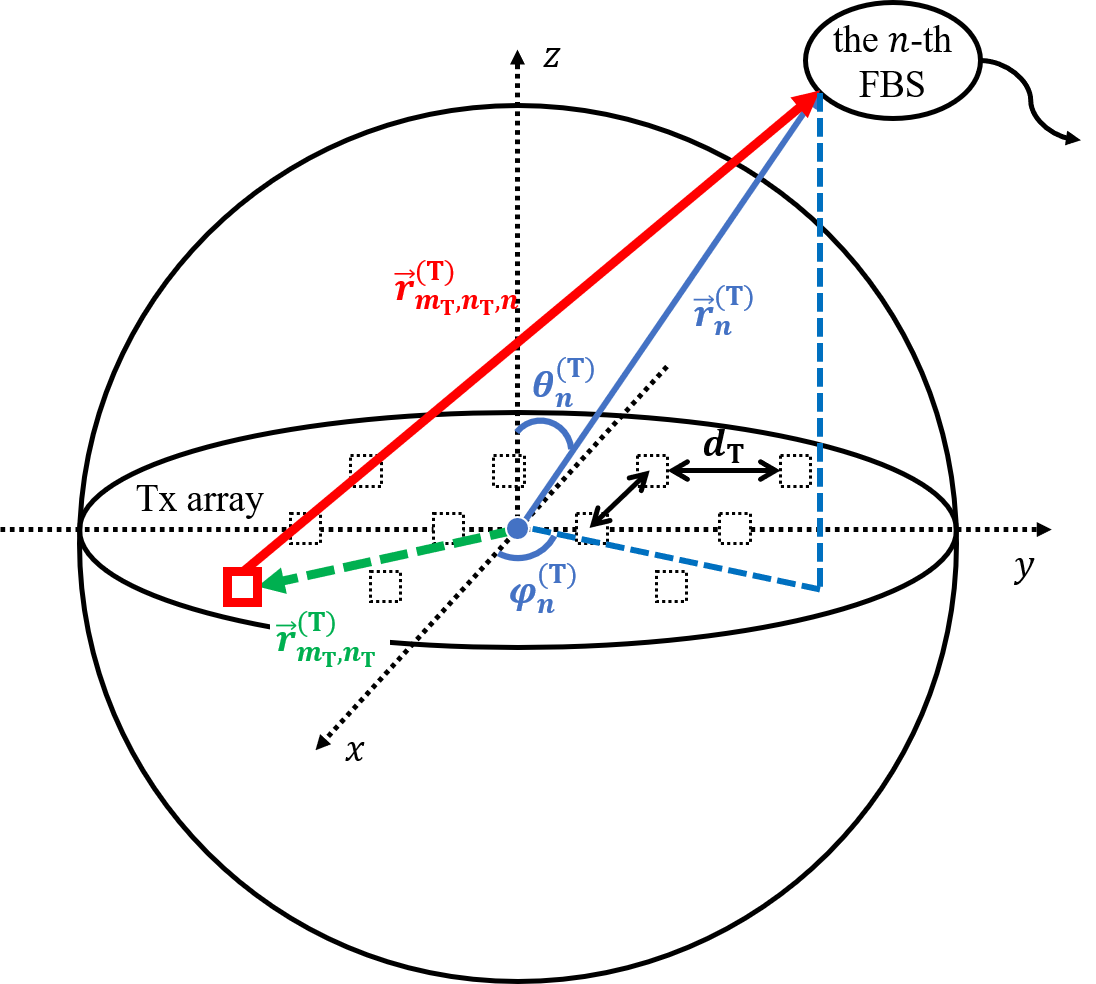}
    \caption{Geometric relation between antenna array center, antenna element, and scatterer (cluster).}
    \label{fig:geometric_center_element_cluster}
\end{figure}

\par The vector from the Tx-UPA center to the FBS is represented by $\vec{r}_{n}^{\rm(T)}$. In the local coordinate system shown in Fig.~\ref{fig:geometric_center_element_cluster}, where the UPA lies in the $xy$-plane and the center of UPA is the origin, ($r_{n}^{\rm(T)}$, $\theta_{n}^{\rm(T)}$, $\varphi_{n}^{\rm(T)}$) indicates the spherical coordinate of the FBS. Note that $\theta_{n}^{\rm(T)}$ is the zenith angle rather than the elevation angle. Denote the $1\times3$ vector $\boldsymbol{r}_{n}^{\rm(T)} = [r_{n}^{\rm(T)} \sin\theta_{n}^{\rm(T)} \cos\varphi_{n}^{\rm(T)}, r_{n}^{\rm(T)} \sin\theta_{n}^{\rm(T)} \sin\varphi_{n}^{\rm(T)}, r_{n}^{\rm(T)} \cos\theta_{n}^{\rm(T)}]$ as
\begin{subequations}
\begin{align}
    \boldsymbol{r}_{n}^{\rm(T)} &= r_{n}^{\rm(T)} \boldsymbol{u}_{n}^{\rm(T)} \\
    \boldsymbol{u}_{n}^{\rm(T)} &= [\sin\theta_{n}^{\rm(T)} \cos\varphi_{n}^{\rm(T)}, \sin\theta_{n}^{\rm(T)} \sin\varphi_{n}^{\rm(T)}, \cos\theta_{n}^{\rm(T)}]
\end{align}    
\end{subequations}
where $\boldsymbol{u}_{n}^{\rm(T)}$ is a unit vector.
\par The vector from the Tx-UPA center to the ($m_{\rm T}$, $n_{\rm T}$)$^\mathrm{th}$ antenna element is represented by $\vec{r}_{m_{\rm T},n_{\rm T}}^{\rm(T)}$. Since the element spacing is $d_{\rm T}$, the Cartesian coordinate of the element in the local coordinate system is ($\delta_{m_{\rm T}} \cdot d_{\rm T}$, $\delta_{m_{\rm T}} \cdot d_{\rm T}$, 0), where $\delta_{m_{\rm T}} = m_{\rm T}-(M_{\rm T}+1)/2$ and $\delta_{n_{\rm T}} = n_{\rm T}-(N_{\rm T}+1)/2$. We denote the $3\times1$ vector 
\begin{equation}
    \boldsymbol{r}_{m_{\rm T},n_{\rm T}}^{\rm(T)} = \left[ \delta_{m_{\rm T}} \cdot d_{\rm T}, \delta_{n_{\rm T}} \cdot d_{\rm T}, 0 \right]^T
\end{equation}
\par The vector from the ($m_{\rm T}$, $n_{\rm T}$)$^\mathrm{th}$ antenna element to the FBS is represented by $\vec{r}_{m_{\rm T},n_{\rm T},n}^{\rm(T)}$, which can be derived by $\vec{r}_{m_{\rm T},n_{\rm T},n}^{\rm(T)} = \vec{r}_{n}^{\rm(T)} - \vec{r}_{m_{\rm T},n_{\rm T}}^{\rm(T)}$. Then the accurate distance between the ($m_{\rm T}$, $n_{\rm T}$)$^\mathrm{th}$ antenna element and the $n$$^\mathrm{th}$ FBS is expressed as
\begin{equation}
\begin{aligned}
    &r_{m_{\rm T},n_{\rm T},n}^{\rm(T)} = \left\|\vec{r}_{m_{\rm T},n_{\rm T},n}^{\rm(T)}\right\| \\
    &= \sqrt{ \left(r_{n}^{\rm(T)}\right)^2 - 2\cdot r_{n}^{\rm(T)} \cdot \boldsymbol{u}_{n}^{\rm(T)} \cdot \boldsymbol{r}_{m_{\rm T},n_{\rm T}}^{\rm(T)} + \left| \boldsymbol{r}_{m_{\rm T},n_{\rm T}}^{\rm(T)} \right|^2 }
\end{aligned}
\end{equation}
which is based on the spherical wave model. For a $M_{\rm T} \times N_{\rm T}$ UPA with element spacing $d_{\rm T}$, the components in the formula can be specified as
\begin{equation}
    \begin{aligned}
        \boldsymbol{u}_{n}^{\rm(T)} \cdot \boldsymbol{r}_{m_{\rm T},n_{\rm T}}^{\rm(T)} =& \sin\theta_{n}^{\rm(T)} \cos\varphi_{n}^{\rm(T)} \cdot \delta_{m_{\rm T}} d_{\rm T} \\
        &+ \sin\theta_{n}^{\rm(T)} \sin\varphi_{n}^{\rm(T)} \cdot \delta_{n_{\rm T}} d_{\rm T}\\
    \end{aligned}
\end{equation}
\begin{equation}
    \left| \boldsymbol{r}_{m_{\rm T},n_{\rm T}}^{\rm(T)} \right|^2 = \left(\delta_{m_{\rm T}} d_{\rm T}\right)^2 + \left(\delta_{n_{\rm T}} d_{\rm T}\right)^2
\end{equation}
For a $1 \times N_{\rm T}$ ULA along the $y$-axis with element spacing $d_{\rm T}$, the components in the formula can be specified as
\begin{equation}
    \boldsymbol{u}_{n}^{\rm(T)} \cdot \boldsymbol{r}_{m_{\rm T},n_{\rm T}}^{\rm(T)} = \sin\theta_{n}^{\rm(T)} \sin\varphi_{n}^{\rm(T)} \cdot \delta_{n_{\rm T}} d_{\rm T}
\end{equation}
\begin{equation}
    \left| \boldsymbol{r}_{m_{\rm T},n_{\rm T}}^{\rm(T)} \right|^2 = \left(\delta_{n_{\rm T}} d_{\rm T}\right)^2
\end{equation}

\subsubsection{Taylor approximation}
\par By two-variable Taylor expansion in terms of ($\delta_{m_{\rm T}} d_{\rm T}$, $\delta_{n_{\rm T}} d_{\rm T}$), the approximation can be derived as follows.
\par To start with, the $1^{\rm st}$ Taylor approximation (planar wave assumption) is the first two terms of Taylor expansion, whose general form is 
\begin{equation} \label{eq:approx_1st_taylor}
    r_{m_{\rm T},n_{\rm T},n}^{\rm(T)} \approx r_{n}^{\rm(T)} - \boldsymbol{u}_{n}^{\rm(T)} \cdot \boldsymbol{r}_{m_{\rm T},n_{\rm T}}^{\rm(T)}
\end{equation}
In this case, the deviation comes from the third term of the Taylor expansion, as
\begin{equation}
    \text{Phase Deviation } = \frac{2\pi}{\lambda} \cdot \frac{\left| \boldsymbol{r}_{m_{\rm T},n_{\rm T}}^{\rm(T)} \right|^2 - \left( \boldsymbol{u}_{n}^{\rm(T)} \cdot \boldsymbol{r}_{m_{\rm T},n_{\rm T}}^{\rm(T)} \right)^2}{2 \cdot r_{n}^{\rm(T)}}
\end{equation}

\par The $2^{\rm nd}$ Taylor approximation is the first three terms of Taylor expansion, whose general form is 
\begin{equation} \label{eq:approx_2nd_taylor}
    r_{m_{\rm T},n_{\rm T},n}^{\rm(T)} \approx r_{n}^{\rm(T)} - \boldsymbol{u}_{n}^{\rm(T)} \cdot \boldsymbol{r}_{m_{\rm T},n_{\rm T}}^{\rm(T)} + \frac{\left| \boldsymbol{r}_{m_{\rm T},n_{\rm T}}^{\rm(T)} \right|^2 - \left( \boldsymbol{u}_{n}^{\rm(T)} \cdot \boldsymbol{r}_{m_{\rm T},n_{\rm T}}^{\rm(T)} \right)^2}{2 \cdot r_{n}^{\rm(T)}}
\end{equation}
In this case, the deviation comes from the fourth term of the Taylor expansion, as
\begin{equation}
    \text{Phase Deviation } = \frac{2\pi}{\lambda} \cdot \frac{ \boldsymbol{u}_{n}^{\rm(T)} \cdot \boldsymbol{r}_{m_{\rm T},n_{\rm T}}^{\rm(T)} - \left( \boldsymbol{u}_{n}^{\rm(T)} \cdot \boldsymbol{r}_{m_{\rm T},n_{\rm T}}^{\rm(T)} \right)^3 }{ 2 \cdot \left(r_{n}^{\rm(T)}\right)^2 }
\end{equation}

\par Taking the ULA along $y$-axis as example, by restricting the phase deviation to $\pi/8$, the $1^{\rm st}$ FF approximation is valid when 
\begin{equation} \label{eq:boundary_1st_taylor_general}
    r_{n}^{\rm(T)} \geq \frac{2D^2\cos^2\theta_{n}^{\rm(T)}}{\lambda}
\end{equation}
The $2^{\rm nd}$ FF approximation is valid when 
\begin{equation} \label{eq:boundary_2nd_taylor_general}
    r_{n}^{\rm(T)} \geq \sqrt{\frac{D^3}{\lambda}\cdot\left(\sin\theta_{n}^{\rm(T)}\sin\varphi_{n}^{\rm(T)}-\sin^3\theta_{n}^{\rm(T)}\sin^3\varphi_{n}^{\rm(T)}\right)}
\end{equation}
where $D=(N_{\rm T}-1) \cdot d_{\rm T}$ is the antenna array size. Among different pairs of angles, the worst case is
\begin{equation} \label{eq:boundary_1st_taylor_worst}
    r_{n}^{\rm(T)} \geq \frac{2D^2}{\lambda} \quad \text{(Rayleigh/Fraunhofer distance)}
\end{equation}
for the $1^{\rm st}$ FF approximation and
\begin{equation} \label{eq:boundary_2nd_taylor_worst}
    r_{n}^{\rm(T)} \geq 0.62\cdot\sqrt{\frac{D^3}{\lambda}}  \quad \text{(Fresnel distance)}
\end{equation}
for the $2^{\rm nd}$ FF approximation.
\par For the UPA case, $D = d_{\rm T}\cdot\sqrt{(M_{\rm T}-1)^2+(N_{\rm T}-1)^2}$. For the UCA case, $D$ is the diameter of the circle.

\subsubsection{Sub-array approximation}
\begin{figure}
    \centering
    \includegraphics[width=0.7\linewidth]{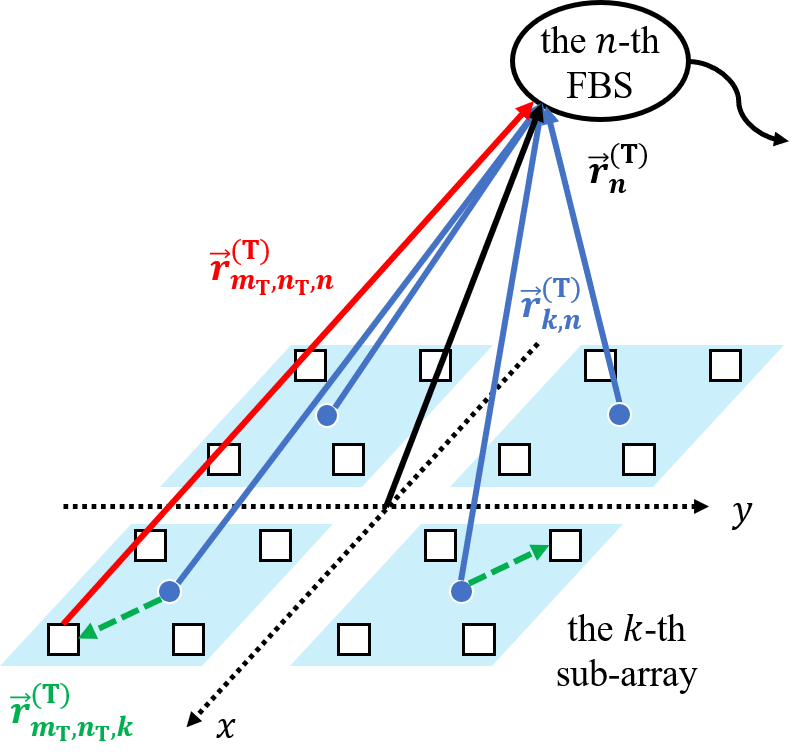}
    \caption{Sub-array division method.}
    \label{fig:sub_array}
\end{figure}
\par Aside from the Taylor approximation, another method is to divide the antenna array into sub-arrays, so that planar wave assumption is valid inside each sub-array~\cite{chen2010hybrid,cui2021near}. Specifically, first, the center of each sub-array is accurately determined by the geometric relation. Then, taking the sub-array center as the reference, the path length from the scatterer to elements in the sub-array is approximated by planar wave assumption.
Detailed expressions are as follows. Assume $K_{m}^{\rm(T)}$ and $K_{n}^{\rm(T)}$ denote the number of sub-arrays in two directions, then each sub-array contains $M_{\rm T}/K_{m}^{\rm(T)} \cdot N_{\rm T}/K_{n}^{\rm(T)}$ elements. First, for the $k$$^\mathrm{th}$ subarray ($k=1,2,...,K_{m}^{\rm(T)}K_{n}^{\rm(T)}$), the vector from the sub-array center to the FBS, $\boldsymbol{r}_{k,n} = r_{k,n}\cdot\boldsymbol{u}_{k,n}$, is accurately calculated. Then, for each element in the  $k$$^\mathrm{th}$ sub-array, vector $\boldsymbol{r}_{m_{\rm T},n_{\rm T},k}^{\rm(T)}$ represents the position of the ($m_{\rm T}$, $n_{\rm T}$)$^\mathrm{th}$ element relative to the Tx-UPA center. Finally, the $1^{\rm st}$ Taylor approximation $r_{m_{\rm T},n_{\rm T},n}^{\rm(T)} \approx r_{k,n}^{\rm(T)} - \boldsymbol{u}_{k,n}^{\rm(T)} \cdot \boldsymbol{r}_{m_{\rm T},n_{\rm T},k}^{\rm(T)}$ is used to calculate the distance between the ($m_{\rm T}$, $n_{\rm T}$)$^\mathrm{th}$ element to the FBS.

\subsubsection{Evaluation and comparison}
\par Take a $N_{\rm T}\times1$ ULA along $y$-axis as an example, assume the carrier frequency is 100~GHz, the element spacing is half-wavelength, the number of elements is 256. The FBS position relative to Tx is defined by $\left( r_{n}^{\rm(T)}, \theta_{n}^{\rm(T)}, \varphi_{n}^{\rm(T)} \right)$ = (5~m, $45^\circ$, $0^\circ$). Sub-array number is $K_{n}^{\rm(T)}=4$, with 64 elements in each sub-array. Under this setting,
\begin{itemize}
    \item The NF-FF boundary of the $1^{\rm st}$ Taylor approximation (planar wave assumption), i.e., Rayleigh distance, is 97.5375~m.
    \item The NF-FF boundary of the $2^{\rm nd}$ Taylor approximation, i.e., Fresnel distance, is 2.6778~m, which is far less than that of the $1^{\rm st}$ Taylor approximation.
    \item The NF-FF boundary of the $1^{\rm st}$ Taylor approximation inside each sub-array is 5.9535~m, which is close to the distance from the scatterer to the array center $r_{n}^{\rm(T)}=$~5~m.
\end{itemize}
\par
\begin{figure}
    \centering
    \includegraphics[width=0.9\linewidth]{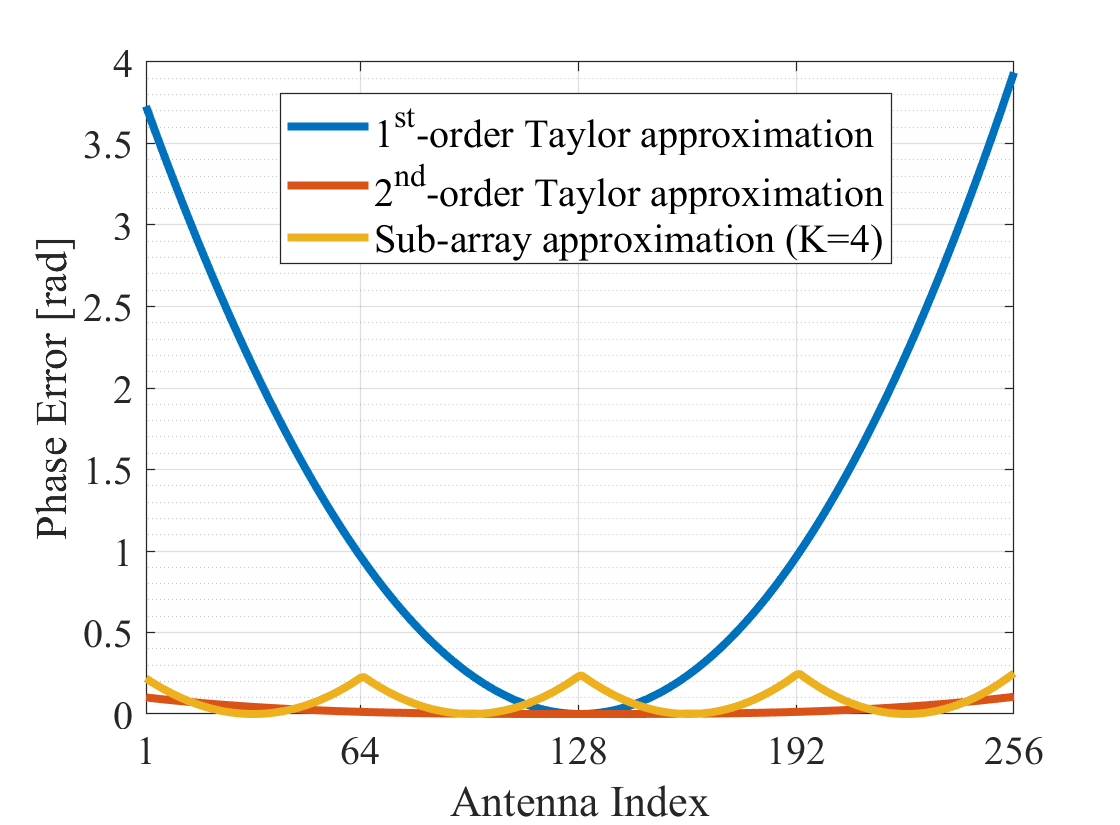}
    \caption{Phase error of different approximation methods.}
    \label{fig:analysis_phase_error}
\end{figure}

 Figure~\ref{fig:analysis_phase_error} shows the comparison of three approximation methods in terms of the phase error at each antenna element. Specifically, using the $1^{\rm st}$ Taylor approximation (planar wave assumption), when the array expands to more than 56 elements, the phase error exceeds $\pi/8=0.3927$ and degrades significantly. Using the sub-array division method, the change of phase error is the same inside each sub-array. However, since the center of each sub-array is accurate, and the division of sub-arrays guarantees the feasibility of planar wave assumption, the maximum phase error is always smaller than $\pi/8$. In contrast, the increment of phase error is obviously slowed down using the $2^{\rm nd}$ Taylor approximation. In this example, the phase error exceeds $\pi/8$ when the number of antenna elements in a ULA reaches 386, which is superfluous in the practical case.

\begin{table*}
    \centering
    \caption{Comparison of the $2^{\rm nd}$ Taylor approximation and sub-array division.}
    \begin{tabular}{|c|c|c|}
    \hline
    \multicolumn{1}{|c|}{\textbf{Method}} & \multicolumn{1}{c|}{\textbf{$2^{\rm nd}$ Taylor approximation}} & \multicolumn{1}{c|}{\textbf{Sub-array division} $\dag$} \\ \hline
    \multicolumn{1}{|c|}{\multirow{2}{*}{Prons}} & \multicolumn{1}{c|}{\multirow{2}{*}{\begin{tabular}[c]{@{}c@{}}direct calculation,\\ free of other steps\end{tabular}}} & \multicolumn{1}{c|}{\begin{tabular}[c]{@{}c@{}}sub-arrays can be reused in the\\ characterization of non-stationarity\end{tabular}} \\ \cline{3-3} 
    \multicolumn{1}{|c|}{} & \multicolumn{1}{c|}{} & \multicolumn{1}{c|}{bounded maximum path length (phase) error} \\ \hline
    \multicolumn{1}{|c|}{Cons} & \multicolumn{1}{c|}{\begin{tabular}[c]{@{}c@{}}unbounded, increasing error with\\ larger antenna array size in theory $\star$ \end{tabular}} & \multicolumn{1}{c|}{\begin{tabular}[c]{@{}c@{}}require addition procedures (and criterion)\\ to divide sub-arrays, high complexity\end{tabular}} \\ \hline
    \multicolumn{3}{l}{\begin{tabular}[c]{@{}l@{}}$\star$ Still, the increment rate of the error is significantly restrained compared with the $1^{\rm st}$ Taylor approximation\\ (planar wave assumption). According to the example above, the number of antenna elements in the ULA needs to\\ reach 386 to exceed the phase error of $\pi/8$, which is superfluous in the practical case.\end{tabular}} \\
\multicolumn{3}{l}{$\dag$ The $1^{\rm st}$ Taylor approximation (planar wave assumption) is used inside each sub-array.}
    \end{tabular}
    \label{tab:approx_compare}
\end{table*}

\subsection{Zenith and Azimuth Angles in Cross-Field}
\par Given the antenna array size $D$ and the position of the scatterer (characterized by the distance between the scatterer and the array center $r_{n}^{\rm(T)}$, zenith angle $\theta_{n}^{\rm(T)}$, and azimuth angle $\varphi_{n}^{\rm(T)}$), we first explore how the actual angle at elements across the array differs from the angle at the array center. We define its maximum difference as the Maximum Angle Difference (MAD).
\par Similar to the definition of Rayleigh distance and Fresnel distance, which derive the NF-FF boundaries of path length by restricting the maximum phase difference to $\pi/8$, we first obtain the express of MAD and then derive the NF-FF boundary of angles by restricting it to half of the half-power beamwidth (HPBW).

\subsubsection{Zenith angle in the ULA case}
\begin{figure}
    \centering
    \begin{subfigure}[3D.]{
    \includegraphics[width=0.45\linewidth]{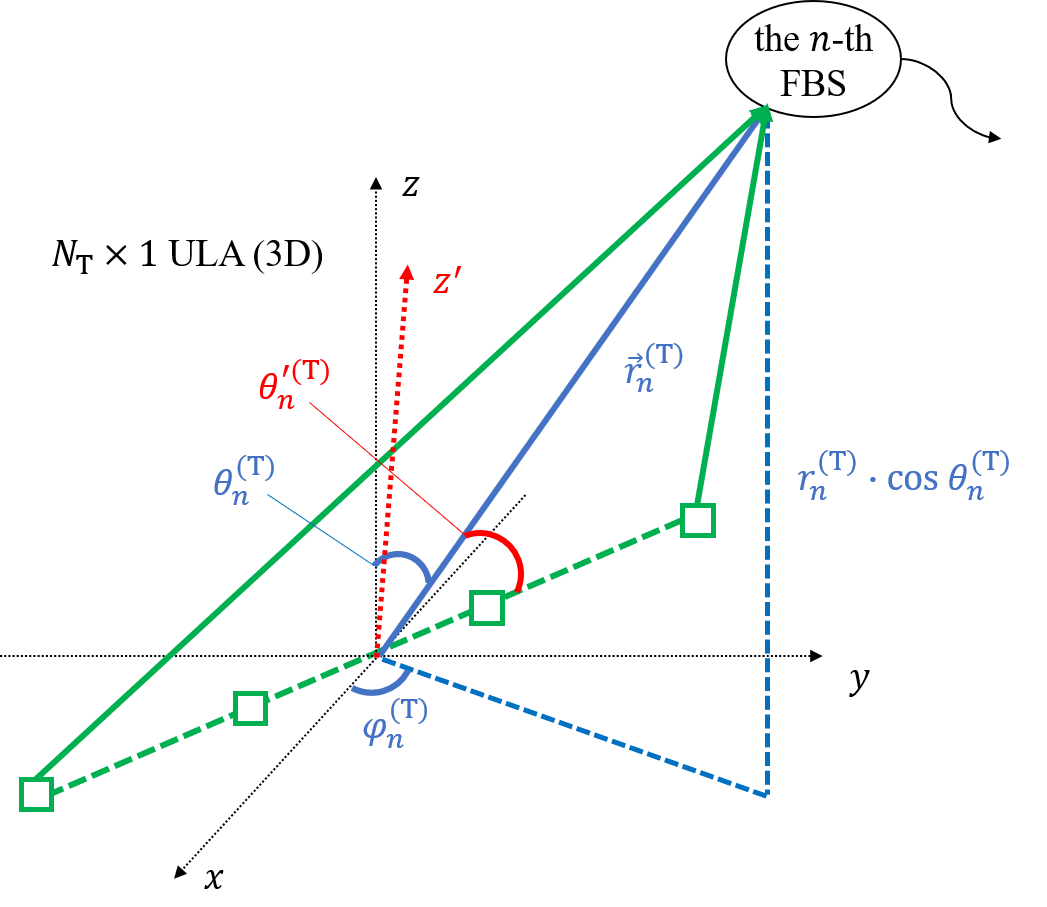}}
    \end{subfigure}
    \begin{subfigure}[2D.]{
    \includegraphics[width=0.45\linewidth]{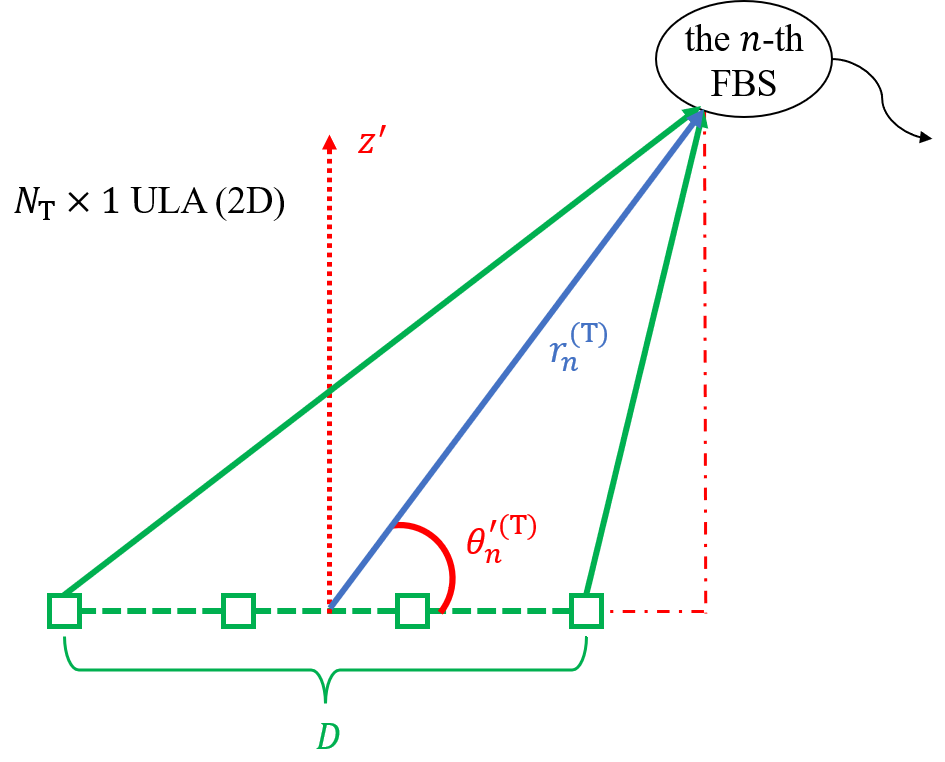}}
    \end{subfigure}
    \caption{Maximum angle difference across the ULA.}
    \label{fig:MAD_ULA}
\end{figure}

\par First, take the $N_{\rm T}\times1$ ULA along the $y$-axis as the example, the expression of MAD is derived from the geometric relation, as
\begin{equation} \label{eq:MAD}
    \text{MAD} = \theta_{n}^{'(T)} - \arctan\frac{(r_{n}^{\rm(T)}/D)\cdot\sin\theta_{n}^{\rm'(T)}-1/2}{(r_{n}^{\rm(T)}/D)\cdot\cos\theta_{n}^{\rm'(T)}} \quad \text{(ULA)}
\end{equation}
Note that since $\arctan(X)<0$ when $x<0$, $\theta_{n}^{\rm'(T)}$ is an elevation angle with the range of $[-90^\circ, 90^\circ]$. In contrast, $\theta_{n}^{\rm(T)}$ in Fig.~\ref{fig:MAD_ULA} is a zenith angle with the range of $[0^\circ, 180^\circ]$. In the same coordinate system, the sum of the elevation angle and the zenith angle is $90^\circ$. However, in the ULA case, $\theta_{n}^{\rm'(T)}$ is not in the same coordinate system with $\theta{n}^{\rm(T)}$ and thus cannot be simply calculated by $90^\circ-\theta_{n}^{\rm(T)}$. Instead, $\theta_{n}^{\rm'(T)}$ is defined by the angle between the ULA and the vector $\vec{r}_{n}^{\rm(T)}$.
\par
\begin{figure}
    \centering
    \includegraphics[width=0.9\linewidth]{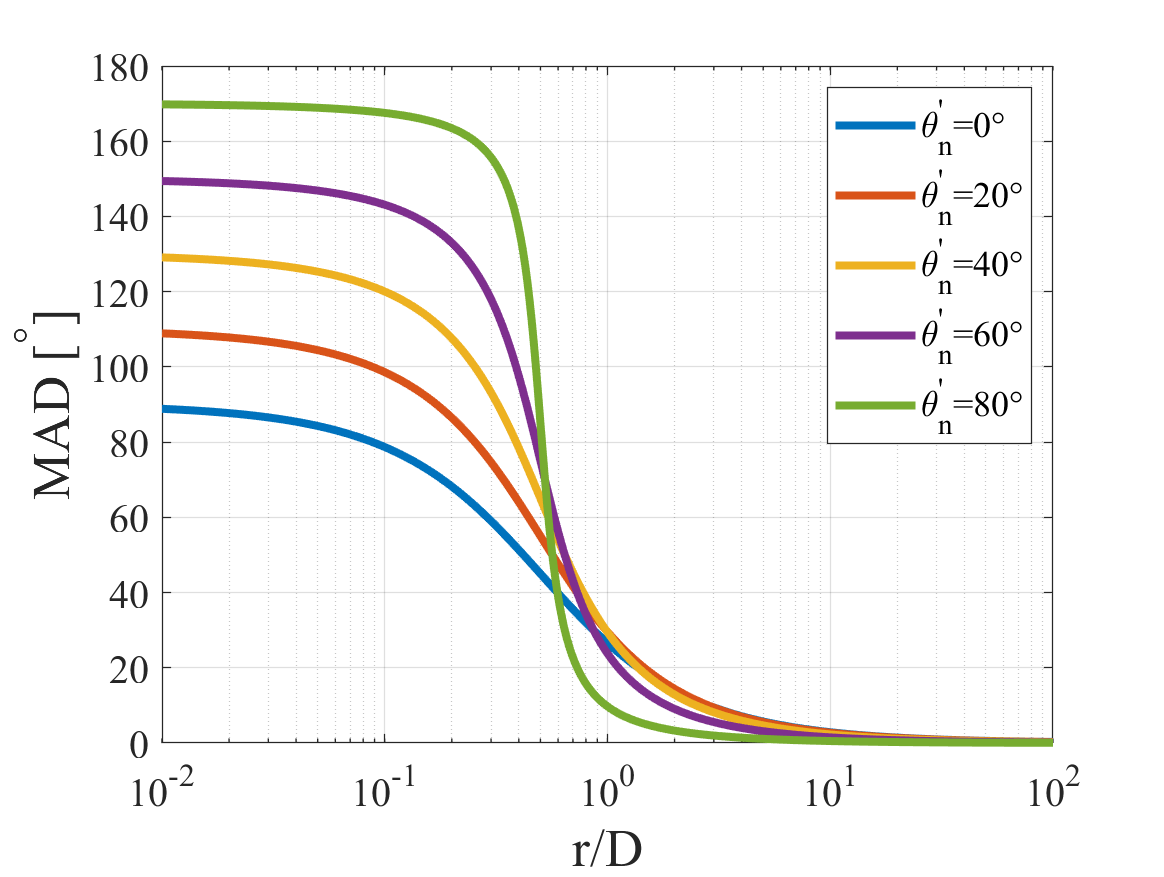}
    \caption{Maximum angle difference (MAD) is monotonically decreasing with the ratio $r_{n}^{\rm(T)}/D$.}
    \label{fig:analysis_mad}
\end{figure}

 When $\theta_{n}^{\rm'(T)}$ is fixed, MAD is monotonically decreasing with the ratio $r_{n}^{\rm(T)}/D$, as shown in Fig.~\ref{fig:analysis_mad}.
By restricting
\begin{equation}
    \text{MAD} < \text{HPBW}_{\rm tx}^{\rm(V)}/2
\end{equation}
where $\text{HPBW}_{\rm tx}^{\rm(V)}$ is the vertical half-power beamwidth, we derive the lower bound of the distance between the scatterer and the array center $r_{n}^{\rm(T)}$ (or the ratio $r_{n}^{\rm(T)}/D$), as
\begin{equation} \label{eq:boundary_zenith_ula}
    r_{n}^{\rm(T)} > \frac{D}{2\cdot\left[ \sin\theta_{n}^{\rm'(T)}-\tan(\theta_{n}^{\rm'(T)}-\text{HPBW}_{\rm tx}^{\rm(V)}/2) \cdot \cos\theta_{n}^{\rm'(T)} \right]}
\end{equation}
which can be regarded as the NF-FF boundary of the zenith angle. If \eqref{eq:boundary_zenith_ula} is valid, we can assume that the zenith angle of the scatter at any antenna element is the same as the zenith angle at the array center, otherwise, the zenith angle should be accurately calculated.

\subsubsection{Zenith angle in UPA and UCA cases}
\begin{figure*}
    \centering
    \begin{subfigure}[Zenith angle difference across UPA.]{
        \includegraphics[width=0.35\linewidth]{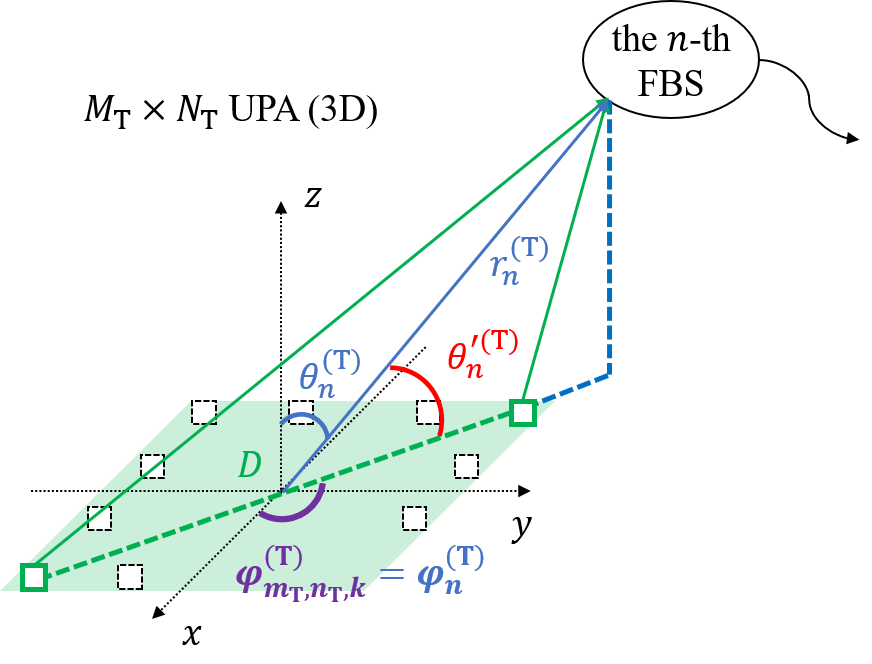}}
    \end{subfigure}
    \begin{subfigure}[Zenith angle difference across UCA.]{
        \includegraphics[width=0.3\linewidth]{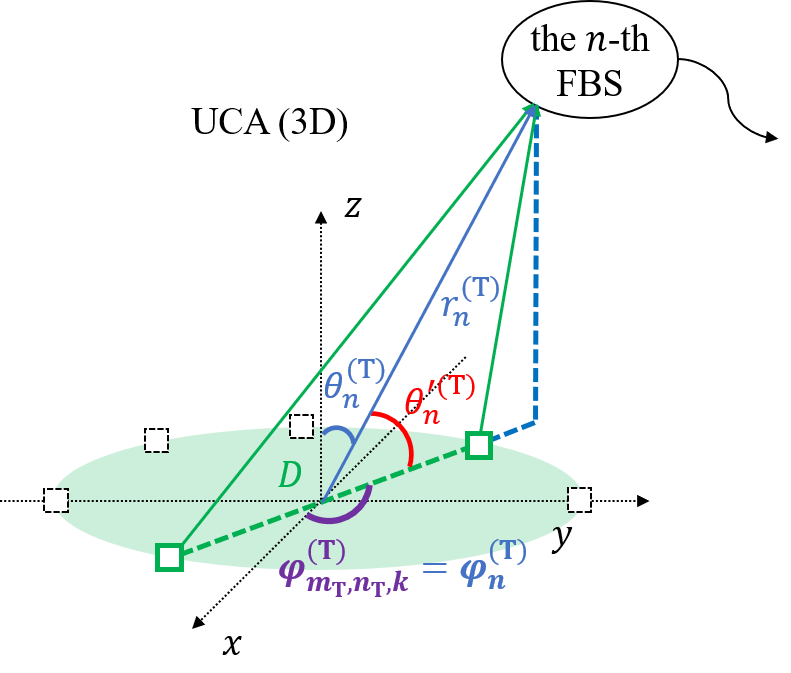}}
    \end{subfigure}
    \begin{subfigure}[Azimuth angle difference.]{
        \includegraphics[width=0.25\linewidth]{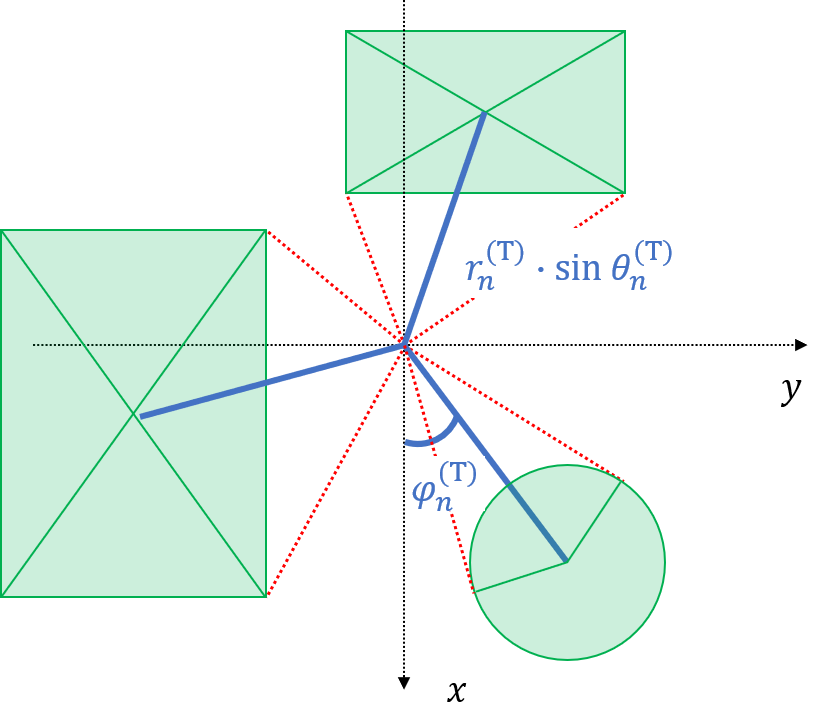}}
    \end{subfigure}
    \caption{Maximum angle difference across UPA and UCA which lie in the $xy$-plane.}
    \label{fig:MAD_UPA_UCA}
\end{figure*}
\par The aforementioned derivation is based on the ULA. In the case of UPA, as shown in Fig.~\ref{fig:MAD_UPA_UCA}(a), the worst case happens when the projection of $\vec{r}_{n}^{\rm(T)}$ coincides with the vector from the array center to the farest element, i.e.,
\begin{equation}
    \exists_{m_{\rm T}\in(1,M_{\rm T}) \atop n_{\rm T}\in(1,N_{\rm T})} \varphi_{m_{\rm T},n_{\rm T},n}^{\rm(T)} = \varphi_{n}^{\rm(T)}
\end{equation}
In this case, we can substitute
\begin{equation}
    D = d_{\rm T}\cdot\sqrt{(M_{\rm T}-1)^2+(N_{\rm T}-1)^2}
\end{equation}
into the derivation in the ULA case and reuse \eqref{eq:MAD} and \eqref{eq:boundary_zenith_ula}. Still, note that the definition of $\theta_{n}^{\rm'(T)}$ is defined as the elevation angle under the local coordinate system that regards the UPA as the $xy$-plane (i.e., the normal vector of the UPA is the $z$-axis), and thus can be derived by $\theta_{n}^{\rm'(T)} = 90^\circ - \theta_{n}^{\rm(T)}$. If the UPA does not lie in the $xy$-plane, we need to calculate $\theta_{n}^{\rm'(T)}$ as $90^\circ$ minus the angle between $\vec{r}_{n}^{\rm(T)}$ and the normal vector of the UPA.
\par Similarly, in the case of UCA, as shown in Fig.~\ref{fig:MAD_UPA_UCA}(b), we can substitute
\begin{equation}
    D=2R
\end{equation}
into \eqref{eq:MAD} and \eqref{eq:boundary_zenith_ula}, where $R$ is the radius of the UCA. Still, the definition of $\theta_{n}^{\rm'(T)}$ is defined as the elevation angle under the local coordinate system that regards the UCA as the $xy$-plane, i.e., the normal vector of the UCA is the $z$-axis. If the UCA does not lie in the $xy$-plane, we need to calculate $\theta_{n}^{\rm'(T)}$ as $90^\circ$ minus the angle between $\vec{r}_{n}^{\rm(T)}$ and the normal vector of the UCA.

\subsubsection{Azimuth angle}
\par The aforementioned derivation is targeted at the zenith/elevation angle. For the azimuth angle, we can observe the projection of the scatterer on the $xy$-plane as shown in Fig.~\ref{fig:MAD_UPA_UCA}(c). The maximum azimuth angle difference occurs at four vertex of the UPA. Therefore, we can simply calculate the azimuth angle difference at these four positions and take their maximum, as
\begin{equation}
\begin{aligned}
    &\text{MAD (azimuth)} = \\
    &\max_{m_{\rm T}\in\{1,M_{\rm T}\} \atop n_{\rm T}\in\{1,N_{\rm T}\}} \left| \varphi_{n}^{\rm(T)} - \arctan\frac{ r_{n}^{\rm(T)} \sin\theta_{n}^{\rm(T)} \sin\varphi_{n}^{\rm(T)} - \delta_{n_{\rm T}} \cdot d_{\rm T} }{ r_{n}^{\rm(T)} \sin\theta_{n}^{\rm(T)} \cos\varphi_{n}^{\rm(T)} - \delta_{m_{\rm T}} \cdot d_{\rm T} } \right|
\end{aligned}
\end{equation}
\par In the case of UCA, as shown in Fig.~\ref{fig:MAD_UPA_UCA}(c), the maximum azimuth angle difference is half of the angle between two tangent lines of the UCA with radius $R$, i.e., $\text{MAD (azimuth)}=\arcsin\left(R/(r_{n}^{\rm(T)} \sin\theta_{n}^{\rm(T)})\right)$.
\par The value of MAD is then compared with half of the horizontal HPBW.
If
\begin{equation} \label{eq:boundary_horizontal}
    \text{MAD (azimuth)} < \text{HPBW}_{\rm tx}^{\rm(H)}/2
\end{equation}
we can assume that the azimuth angle of the scatterer at any antenna element is the same as that at the array cente. Otherwise, the azimuth angle should be accurately calculated. 

\subsection{Path Gain in Cross-Field}
\par While the antenna gain difference is larger than 3~dB across the array resulting from the deviation of incident angles that exceeds half of the half-power beamwidth (HPBW), another smaller-scale path gain deviation that should be involved in cross-field channel modeling is the gain reduction factor of the waveguides in the near-field~\cite{wang2024far,xiao2021near}.
In specific, the free-space path loss of the vertically incident ray at each element can be statistically modeled by
\begin{subequations}
\begin{align}
    &\text{PL [dB]} = 20\times\log_{10}\frac{4\pi}{\lambda}\cdot d_{\rm ref}\cdot K(\Delta x,\Delta y,\lambda,d_{0}), \\
    &K(\Delta x,\Delta y,\lambda,d_{0}) = 1 + C_{1}^{\frac{\left(\Delta x / C_3\right)^2 + \left(\Delta y / C_4\right)^2}{\lambda} - C_2\cdot d_{0} },
\end{align}
\label{eq:gain}
\end{subequations}
where $K$ is the cross-field factor associated with the wavelength $\lambda$ (or equivalently the frequency $f$), the distance from the scatter to the perpendicular array, and the coordinate of the element relative to the center ($\Delta x$, $\Delta y$). The constant reference distance $d_{\rm ref}$ is a model parameter.
In the proposed cross-field MIMO channel model, it can applied to the LoS ray when antenna arrays are face-to-face. Though, the significance of this gain variation is so far inconclusive, and more efforts are required to generalize the model to arbitrary postures of antenna arrays.

\subsection{Evaluation and Discussions}
\par In the near-field channel modeling, for each parameter of the cluster, i.e., path length (or delay) and angles, we use the NF-FF boundary to determine whether the corresponding approximation method of the parameter is valid. The approximation of the parameter is valid in the FF field, otherwise, we require a more accurate calculation of the parameter.
The following table summarizes the NF-FF boundaries corresponding to the approximation method of small-scale parameters.

\begin{table*}
    \centering
    \caption{Approximation method of path length and angles, and corresponding NF-FF boundaries.}
    \begin{tabular}{|c|c|c|c|}
         \hline
         \textbf{Parameter} & \textbf{FF approximation} & \textbf{Max deviation w.r.t array center} & \makecell{\textbf{NF-FF boundary} \\ \textbf{(BS-FBS as the example)}} \\ \hline
         Path length & $1^{\rm st}$ approx. \eqref{eq:approx_1st_taylor}  & \makecell{$\lambda/16$ in path length, \\ or $\pi/8$ in phase} & general: \eqref{eq:boundary_1st_taylor_general}; worst: \eqref{eq:boundary_1st_taylor_worst} \\ \hline
         Path length & $2^{\rm nd}$ approx. \eqref{eq:approx_2nd_taylor}  & \makecell{$\lambda/16$ in path length, \\ or $\pi/8$ in phase} & general: \eqref{eq:boundary_2nd_taylor_general}; worst: \eqref{eq:boundary_2nd_taylor_worst} \\ \hline
         Zenith angle & same as that at array center & $\text{HPBW}_{\rm tx}^{\rm(V)}/2$ & \eqref{eq:boundary_zenith_ula} \\ \hline
         Horizontal angle & same as that at array center & $\text{HPBW}_{\rm tx}^{\rm(H)}/2$ & \eqref{eq:boundary_horizontal} \\ \hline
    \end{tabular}
    \label{tab:boundary_compare}
\end{table*}
\par Since antennas are usually/nearly omni-directional in the horizontal direction, we can assume the horizontal angle at each element remains the same as that at the array center.
\par
\begin{figure}
    \centering
    \includegraphics[width=\linewidth]{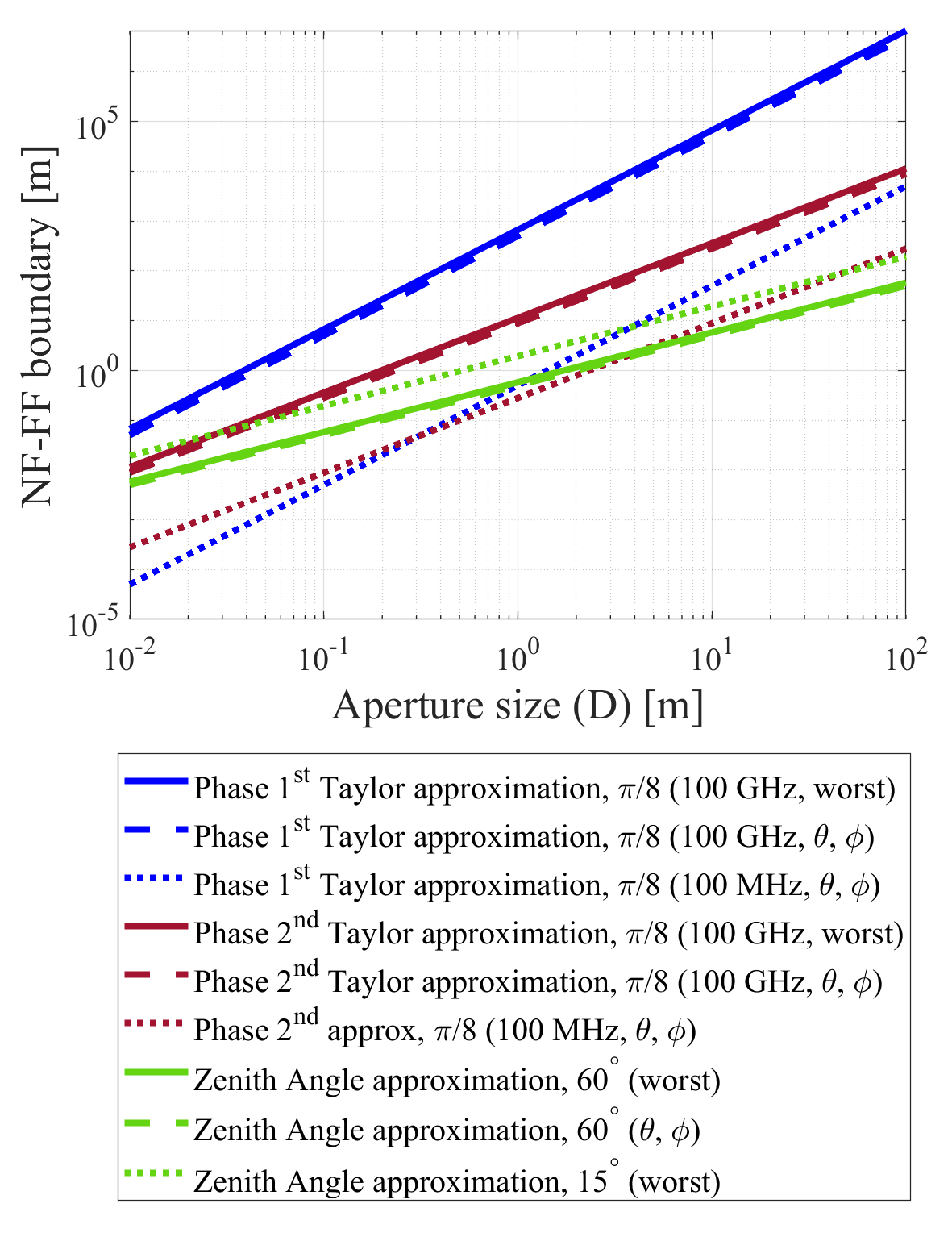}
    \caption{NF-FF boundaries, in terms of distance between the scatterer and the array center ($r_{n}^{\rm(T)}$ or $r_{n}^{\rm(R)}$), under different parameters.}
    \label{fig:analysis_boundary}
\end{figure}

 Take carrier frequency $f=100$~GHz, azimuth angle of departure $\varphi_{n}^{\rm(T)=45^\circ}$, zenith angle of departure $\theta_{n}^{\rm(T)}=45^\circ$, max phase deviation $\pi/8$, vertical HPBW at Tx $\text{HPBW}_{tx}^{\rm(V)}=120^\circ$ (in other words, max zenith angle difference $60^\circ$) as the example, Fig.~\ref{fig:analysis_boundary} compares the NF-FF boundaries, i.e., \eqref{eq:boundary_1st_taylor_general}, \eqref{eq:boundary_2nd_taylor_general}, and \eqref{eq:boundary_zenith_ula}, under different antenna array size $D$.
\par Meanwhile, NF-FF boundaries in the worst case are also depicted in the same figure. The worst/largest NF-FF boundaries of path length, \eqref{eq:boundary_1st_taylor_worst} and \eqref{eq:boundary_2nd_taylor_worst}, are derived by traversing the value of angles ($\theta_{n}^{\rm(T)}$, $\varphi_{n}^{\rm(T)}$), while the worst/largest NF-FF boundary of zenith angle is derived by traversing the value of angles ($\theta_{n}^{\rm(T)}$, $\varphi_{n}^{\rm(T)}$) at $\text{HPBW}_{\rm tx}^{\rm(V)}/2=60^\circ$. For arbitrary $\text{HPBW}_{\rm tx}^{\rm(V)}$, the NF-FF boundary of the zenith angle becomes larger/worse when $\text{HPBW}_{\rm tx}^{\rm(V)}$ gets smaller.
\par As we can observe from the result,
\begin{itemize}
    \item NF-FF boundaries, in terms of distance between the scatterer and the array center ($r_{n}^{\rm(T)}$ or $r_{n}^{\rm(R)}$), is proportional to the power of antenna array size $D$. Specifically, the NF-FF boundary of the zenith angle is proportional to $D$, the NF-FF boundary of path length using the $2^{\rm nd}$ Taylor approximation is proportional to $D^{1.5}$, and the NF-FF boundary of path length using the $1^{\rm st}$ Taylor approximation is proportional to $D^{2}$. As a result, when the antenna array size $D$ grows, the FF approximation is more likely to fail for the path length with and then for the zenith angle.
    \item The NF-FF boundaries of angles are independent of the carrier frequency (or wavelength), while the NF-FF boundaries of path length are relevant to the carrier frequency.
    \item For the NF-FF boundary of path loss, at 100~GHz, the boundary at specific angles (blue/red long dashed lines) is close to the worst boundary (blue/red solid line). In contrast, the order of the carrier frequency directly affects the order of the boundary, which means the frequency is more influential to the boundary than the angles. For instance, when the frequency decreases from 100~GHz to 100~MHz, the order of the boundary decreases correspondingly (blue/red short dashed line).
    \item For the NF-FF boundary of angles, at 100~GHz, the boundary at specific angles (green long dashed lines) is close to the worst boundary (green solid line). The boundary does not change with frequency. However, the value of the HPBW is more influential to the boundary than the value of angles (green short dashed line).
\end{itemize}

\section{Cross-Field Channel Measurement and Result} \label{sec:campaign}

In this section, we describe the measurement conducted in the cross near- and far-field scenario.

\begin{figure}
    \centering
    \includegraphics[width=0.9\linewidth]{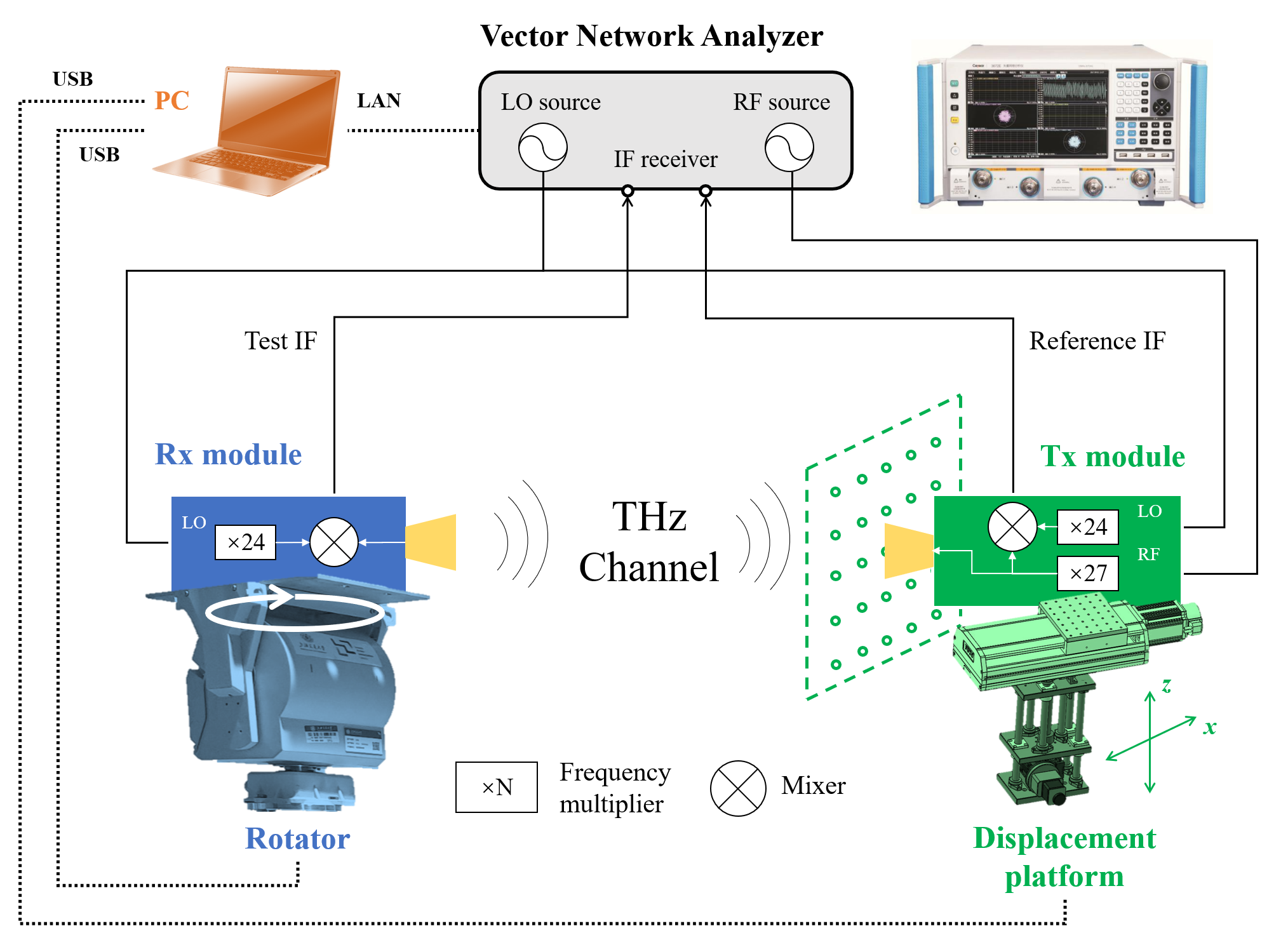}
    \caption{The diagram of the measuring system.}
    \label{fig:platform}
\end{figure}
\begin{figure}
    \centering
    \subfigure[The photo of the measurement.]{
    \includegraphics[width=0.7\linewidth]{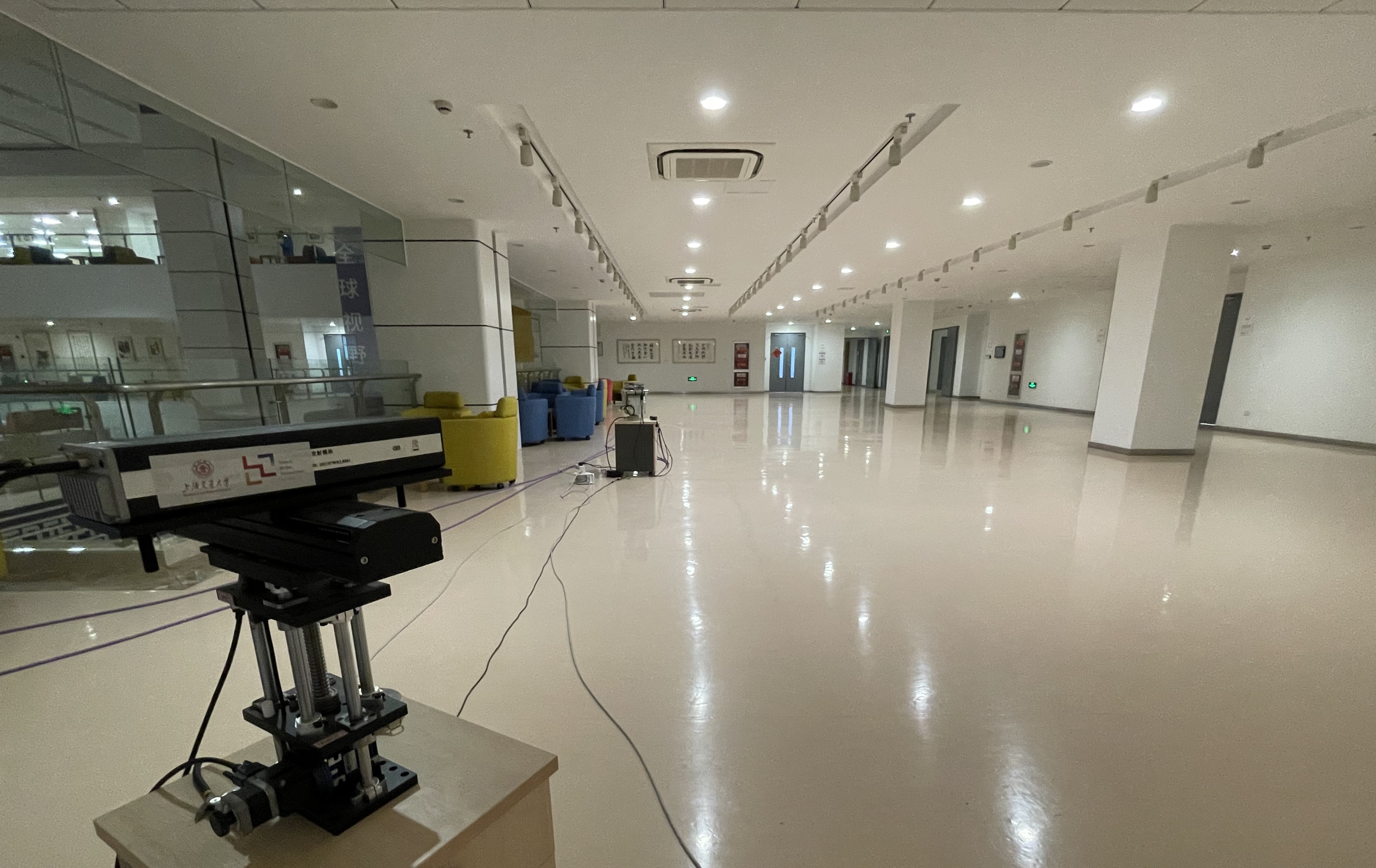}
    }
    \\
    \subfigure[Measurement deployment.]{
    \includegraphics[width=0.7\linewidth]{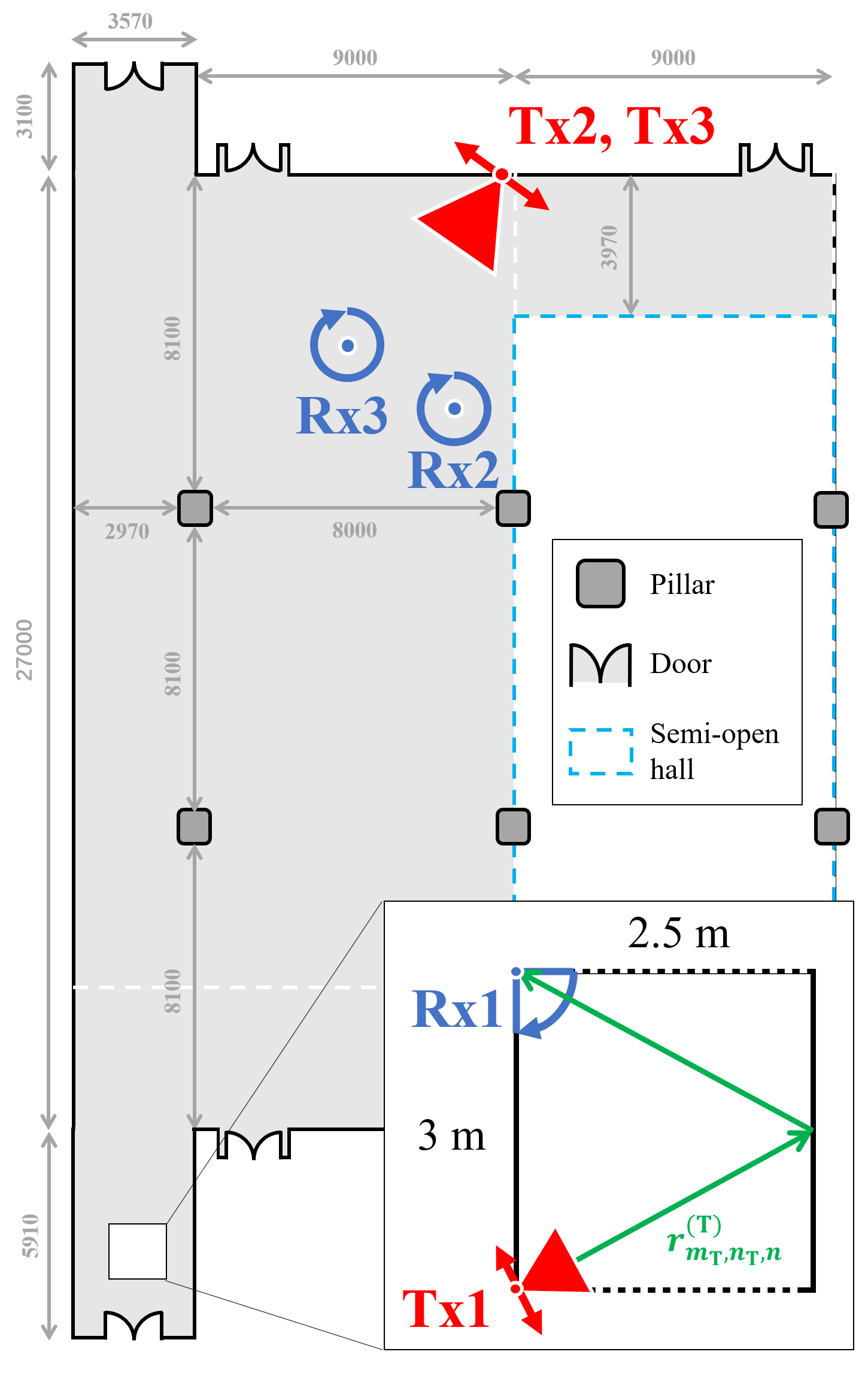}
    }
    \caption{Overview of the measurement.}
    \label{fig:measurement}
\end{figure}

\subsection{Channel Measurement System}
The THz MISO channel measurement system is composed of three parts, as shown in Fig.~\ref{fig:platform}, including the computer (PC) as the control platform, the Ceyear 3672C vector network analyzer (VNA), the displacement platform carrying the THz transmitter (Tx) module, and the rotator carrying the THz receiver (Rx) module.

The measuring part consists of Tx and Rx modules and the VNA.
The VNA generates radio frequency (RF) and local oscillator (LO) sources.
The RF signal is multiplied by 27 to reach the carrier frequencies. The LO signal is multiplied by 24. As designed, the mixed intermediate frequency (IF) signal has a frequency of 7.6~MHz.
Two IF signals at Tx and Rx modules, i.e., the reference IF signal and the test IF signal, are sent back to the VNA, and the transfer function of the device under test (DUT) is calculated as the ratio of the two frequency responses. The DUT contains not only the wireless channel, but also the device, cables, and waveguides. To eliminate their influence, the system calibration procedure is conducted, which is described in detail in our previous works~\cite{wang2022thz, li2022channel}.

The mechanical part is composed of the displacement platform and the rotator. The displacement platform carries the Tx module to move linearly in the $x$-axis (horizontally) and the $z$-axis (vertically). The rotator carries the Rx module to scan the horizontal plane to receive angular-resolved multi-path components (MPCs).
The PC alternately controls the movement of Tx through the displacement platform, the movement of Rx through the rotator, and the measuring process through the VNA.
The measurement starts from the virtual antenna element in Tx at the left bottom corner, and all elements in the UPA are scanned first horizontally (in the $x$-axis) and then vertically (in the $z$-axis). At each element in Tx, the Rx scans horizontally and measures the THz channel once per angle.

\subsection{Measurement Deployment}

\begin{table*}
  \centering
  \caption{Summary of measurement parameters.}
    \setlength{\tabcolsep}{6mm}{
    \begin{tabular}{|>{\raggedright\arraybackslash}p{0.3\linewidth}|c|c|c|}
    \toprule
    Parameter & Measurement 1&\multicolumn{2}{|c|}{Measurement 2}\\
    \midrule
    Frequency band              & \multicolumn{3}{|c|}{260-320~GHz}\\ \hline 
    Bandwidth                   & \multicolumn{3}{|c|}{60~GHz}\\ \hline 
    Time resolution             & \multicolumn{3}{|c|}{16.7~ps}\\ \hline 
    Space resolution            & \multicolumn{3}{|c|}{5~mm}\\ \hline 
    Sweeping interval           & 30~MHz&\multicolumn{2}{|c|}{10~MHz}\\ \hline 
    Sweeping points             & 2001&\multicolumn{2}{|c|}{6001}\\ \hline 
    Maximum excess delay        & 33.3~ns&\multicolumn{2}{|c|}{100~ns}\\ \hline 
    Maximum path length         & 10~m&\multicolumn{2}{|c|}{30~m}\\ \hline 
 Notation& Tx1-Rx1& Tx2-Rx2&Tx3-Rx3\\ \hline 
    Tx gain / HPBW& 25~dBi / 8$^\circ$&\multicolumn{2}{|c|}{7~dBi / 60$^\circ$}\\ \hline 
 Tx antenna array type& 8$\times$8 UPA&\multicolumn{2}{|c|}{2$\times$2 UPA}\\ \hline 
 Tx antenna element spacing& 8~mm&56~mm&48~mm\\ \hline 
 Tx aperture size& \multicolumn{2}{|c|}{0.0792~m}&0.0679~m\\ \hline 
 Rayleigh distance& \multicolumn{2}{|c|}{12.1259~m}&8.9088~m\\ \hline 
 Rx gain / HPBW& \multicolumn{3}{|c|}{25~dBi / 8$^\circ$}\\ \hline 
 Rx angular scan& 0$^\circ$:2$^\circ$:90$^\circ$&\multicolumn{2}{|c|}{0$^\circ$:2$^\circ$:358$^\circ$}\\ \hline
    \end{tabular}
    }
  \label{tab:system_parameter}
\end{table*}

In this measurement, we investigate the THz frequency band ranging from 260~GHz to 320~GHz, which covers a substantially large bandwidth of 60~GHz. As a result, the time and space resolution is 16.7~ps and 5~mm, respectively.
Key parameters of the measurement are summarized in Table~\ref{tab:system_parameter}.

First, a small-scale measurement, from Tx1 to Rx1, is carried out, as shown in Fig.~\ref{fig:measurement}(b). The goal of the measurement is to investigate the cluster parameter from scatterers (the wall, in this deployment) in the near-field region. A virtual UPA with $8\times8$ elements is measured at Tx, with the element spacing of 8~mm. In this case, the aperture size reaches 79.2~mm, which corresponds to the Rayleigh distance of 12.1259~m.
The scale of the scenario is 2.5~m $\times$ 3~m, which guarantees that the reflection happens in the near-field region. Furthermore, the small scale of the scenario enables the reduction of the maximum detectable path length to 10~m, corresponding to the frequency sweeping interval of 30~MHz and sweeping points of 2001 for each channel measurement.
To increase the angular resolution, at Rx, the rotator scans from $0^\circ$ to $90^\circ$ with the step of $2^\circ$. Furthermore, high-directive antennas are integrated at both Tx and Rx modules, with gains of around 25~dBi.

Then, another measurement is carried out in a larger-scale scenario, as shown by Tx2-Rx2 and Tx3-Rx3 in Fig.~\ref{fig:measurement}(a) and (b). Different from the setup in the first measurement, the maximum detectable path length is increased to 30~m, corresponding to the frequency sweeping interval of 10~MHz and sweeping points of 6001 for each channel measurement. Besides, the scan range of arrival angles covers $0^\circ$ to $360^\circ$. Under this circumstance, the measuring time at each antenna element is multiplied by 12. Therefore, in this measurement, we decrease the number of virtual antenna element positions from 8$\times$8 to 2$\times$2, while the scale of the antenna array remains at 79.2~mm (and 67.9~m for comparison). Moreover, another antenna with a larger half-power beamwidth (HPBW), $60^\circ$, is equipped at Tx.

\subsection{Measurement Results}

\subsubsection{Measurement~1}
\begin{figure*}
    \centering
    \begin{subfigure}[Delay.]{
    \includegraphics[width=0.3\linewidth]{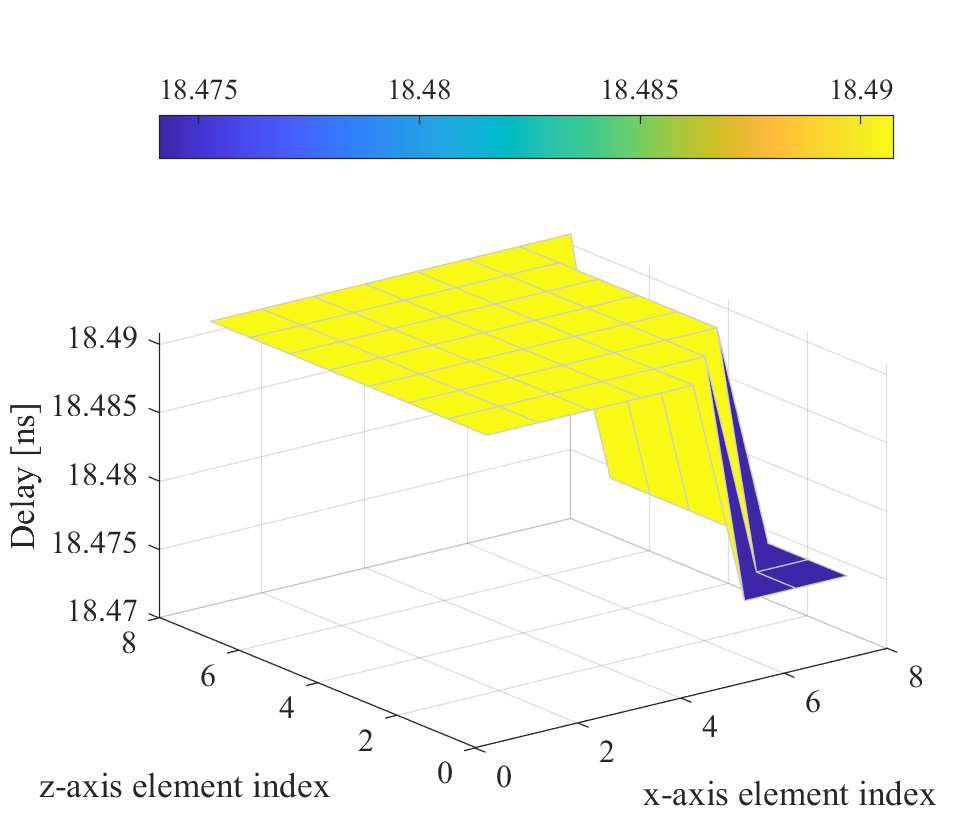}}
    \end{subfigure}
    \begin{subfigure}[Angle of arrival.]{
    \includegraphics[width=0.3\linewidth]{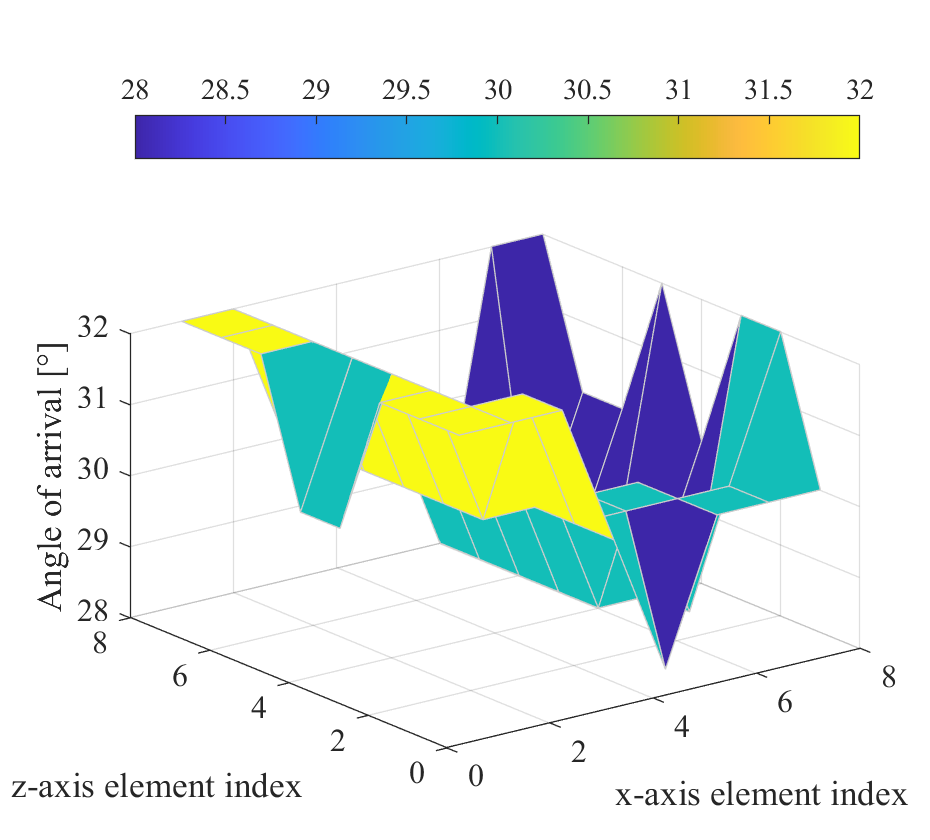}}
    \end{subfigure}
    \begin{subfigure}[Path gain.]{
    \includegraphics[width=0.35\linewidth]{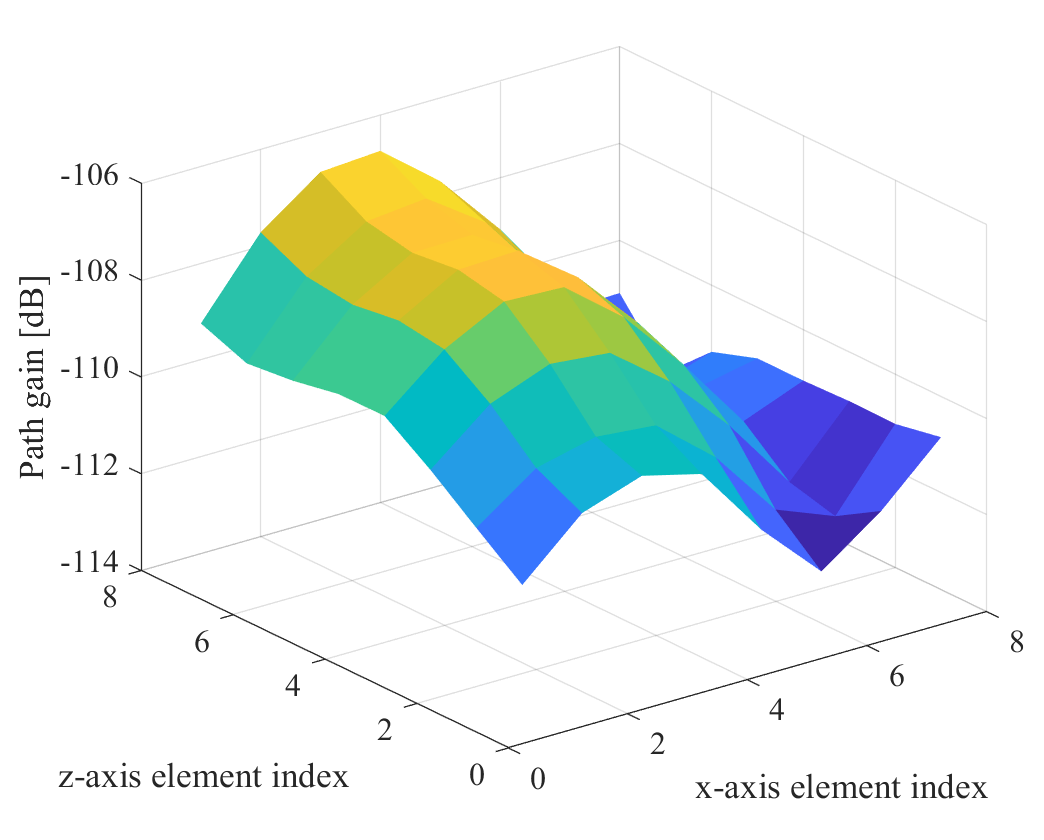}}
    \end{subfigure}
    \caption{Properties of the wall-reflected ray.}
    \label{fig:reflected_ray_properties}
\end{figure*}
The sample with the greatest path gain in the PDAP is regarded as the wall-reflected ray from the Tx to the Rx.
Fig.~\ref{fig:reflected_ray_properties} illustrates the variation, across $8\times8$ elements in the UPA, of delay, arrival angle, and path gain of the wall-reflected ray, respectively.
First, due to symmetry, the distance from each element at Tx to the scatterer is half of the total path length. As shown in Fig.~\ref{fig:reflected_ray_properties}(a), the variation in distance from elements at Tx to the scatterer exceeds 2.5~mm, which is half of the path length resolution. The deviation is larger than $1/16$ of the wavelength, 0.065~mm, which accords with the analysis in the near-field region as summarized in Table~\ref{tab:boundary_compare}.

Second, the theoretical variation in the departure angle is about $1.5^\circ$ across 64 elements at Tx. Due to the reflection, scattering, and influence of antenna radiation patterns, the variation in the arrival angle reaches $4^\circ$ and results in a $8$-dB variation in path gain of the measured MPC, as shown in Fig.~\ref{fig:reflected_ray_properties}(b) and (c). Despite the inevitable superposition of power inside the received beam in the practical measurement, in the simulation of cross-field channels, the departure/arrival angle of the MPC should still be carefully modeled when scatterers are in the near-field region, as the antenna gain differs in the antenna radiation pattern at different angles.

\begin{figure*}
    \centering
    \begin{subfigure}[Cluster delay.]{
    \includegraphics[width=0.31\linewidth]{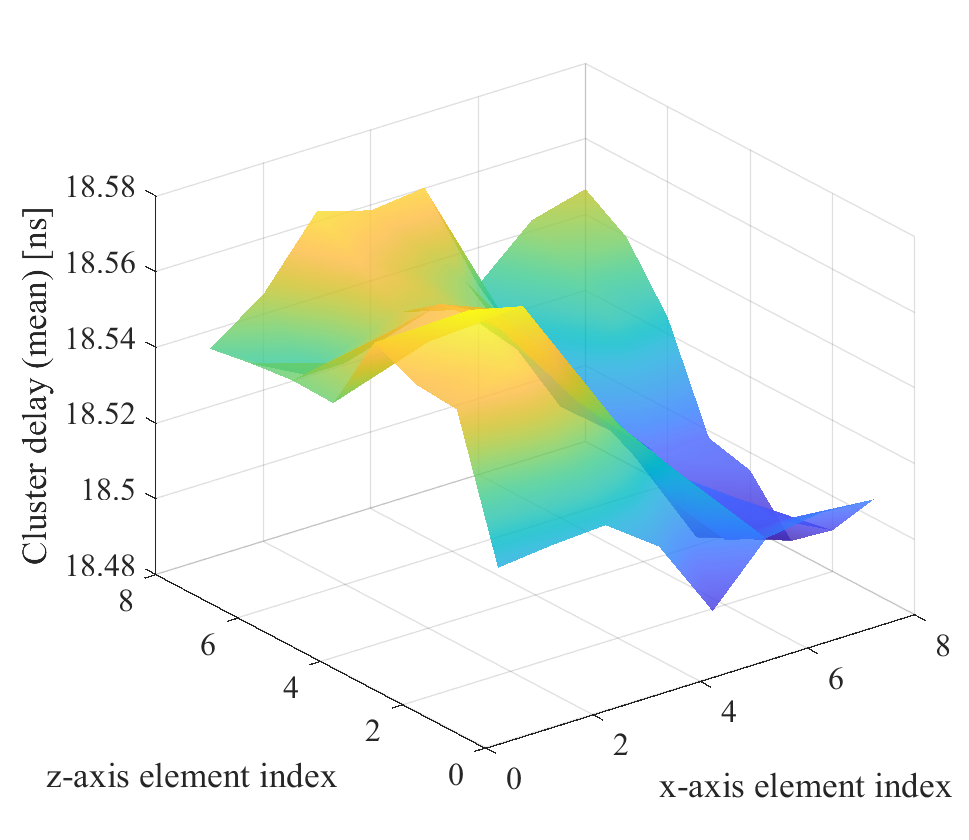}}
    \end{subfigure}
    \begin{subfigure}[Intra-cluster delay spread.]{
    \includegraphics[width=0.31\linewidth]{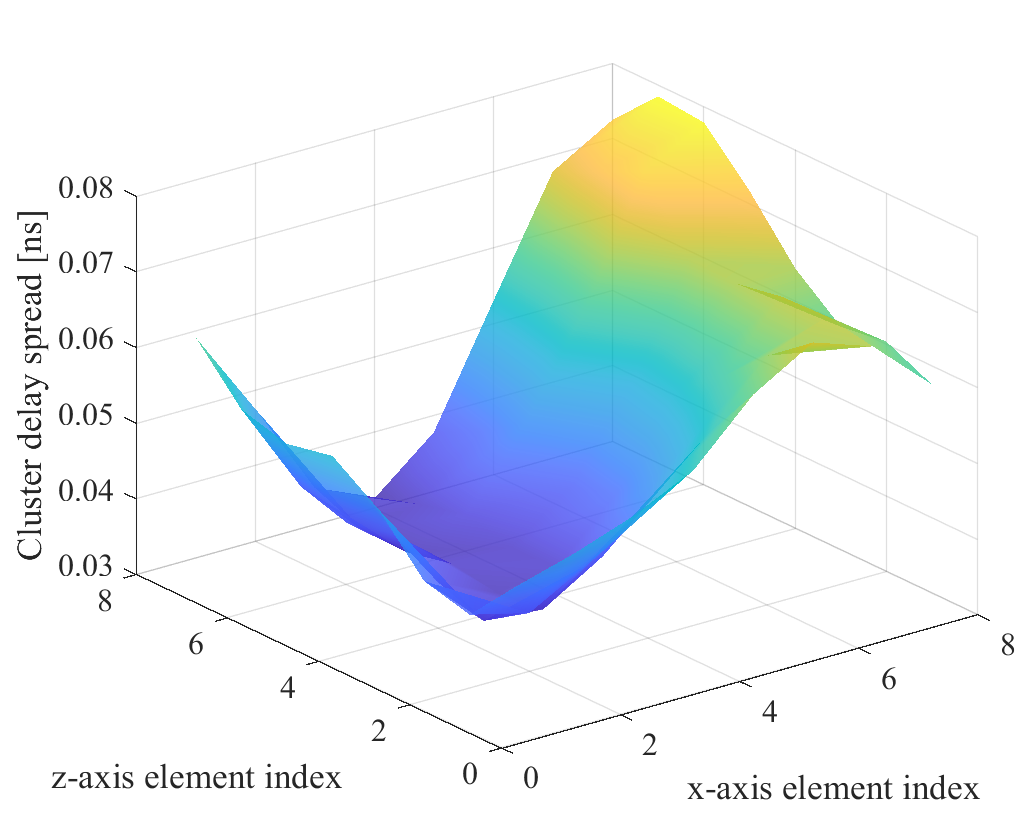}}
    \end{subfigure}
    \begin{subfigure}[Intra-cluster angular spread.]{
    \includegraphics[width=0.31\linewidth]{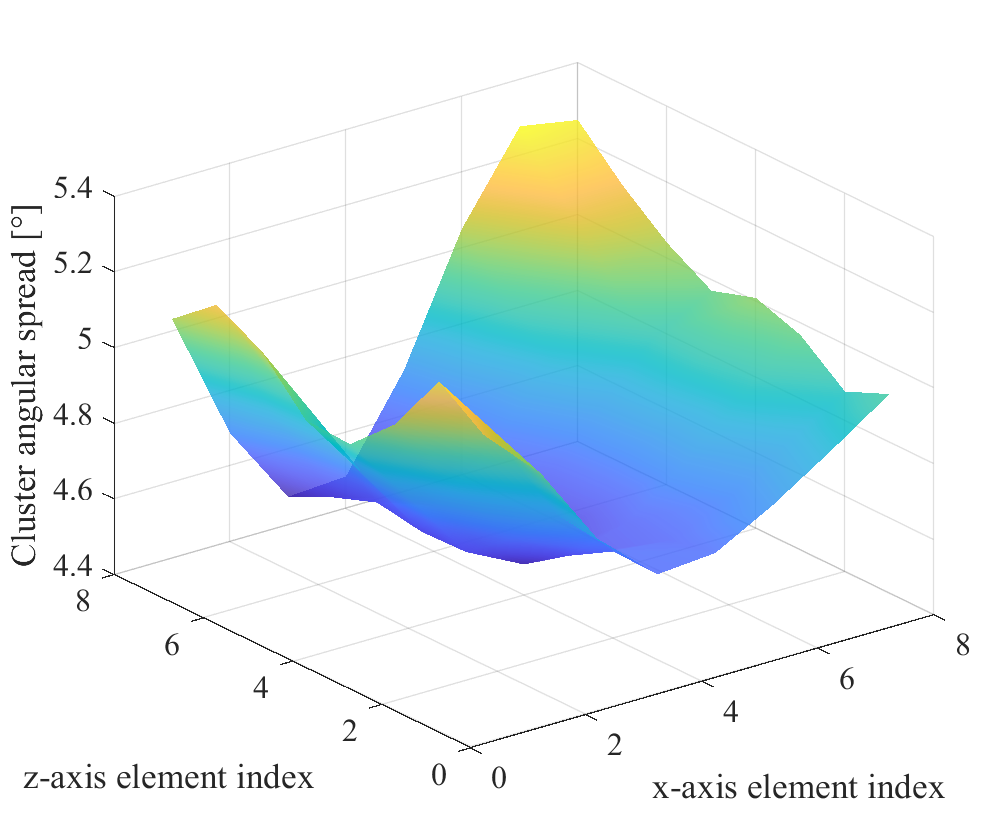}}
    \end{subfigure}
    \caption{Properties of the cluster centered on the wall-reflected ray.}
    \label{fig:reflected_ray_cluster_properties}
\end{figure*}

Furthermore, considering the cluster of MPCs around the wall-reflected ray, cluster parameters also vary across different elements at Tx, as summarized in Fig.~\ref{fig:reflected_ray_cluster_properties}.

\subsubsection{Measurement~2}

In the measurement in a larger indoor scenario, power-delay-angle profiles (PDAPs) are basically the same at four Tx antenna elements for Tx2-Rx2 and Tx3-Rx3 channels, respectively. 8-14 clusters are observed from Tx2 to Rx2, and 9-13 clusters are observed from Tx3 to Rx3, while the difference in the number of clusters comes from the difference in received power and the classification method which distinguishes MPCs and noise samples by thresholding~\cite{wang2022thz}.

The channel from the $s$$^\mathrm{th}$ antenna element at Tx to the Rx ($s=1,2,...,4$) is represented by $\mathbf{h}_{s}$, $1\times6001$ complex gain. We denote $\mathbf{H}=\left[\mathbf{h}_1,\mathbf{h}_2,...,\mathbf{h}_S\right]^{\rm T}$, and calculate the transmitter spatial correlation matrix as
\begin{equation}
    \mathbf{R}_{\rm(T)} = \frac{\mathbf{H}\mathbf{H}^{*}}{\left\|\mathbf{H}\right\| \left\|\mathbf{H}\right\|^{\rm T}}
\end{equation}
where $*$ is the complex conjugate transpose of the matrix.
$\left\|\mathbf{H}\right\| = \left[ \left\|\mathbf{h}_1\right\|_{F}, \left\|\mathbf{h}_2\right\|_{F}, ..., \left\|\mathbf{h}_S\right\|_{F} \right]^{\rm T}$ where $\left\|\mathbf{h}_s\right\|_{F}^2 = \text{tr}(\mathbf{h}_s^{\rm T}\mathbf{h}_s)$ is the Frobenius norm.
Assume there are adequate antennas at the receiver, the effective degrees of freedom (EDoF) is constrained by the transmitter spatial correlation matrix, which is calculated by
\begin{equation}
    \text{EDoF}_{\rm(T)\ constrained} = \left( \frac{ \text{tr}\left( \mathbf{R}_{\rm(T)} \right) }{ \left\|\mathbf{R}_{\rm(T)}\right\|_{F} } \right)^2
\end{equation}
The Rayleigh distance is 12.1~m for the antenna array at Tx2, and the counterpart is 8.9~m for the antenna array at Tx3. According to the PDAP results, more scatterers are located in the NF region for Tx2-Rx2 transmission than for Tx3-Rx3 transmission. Correspondingly, the EDoF constrained by the Tx antenna array is 3.3460 for Tx2, and the counterpart is 3.3031 for Tx3, which are close to 4, the number of antenna elements at Tx. 
These values indicate the low correlation between antenna elements at Tx for near-field and cross-field channels, and the necessity of element-level modeling of channels with scatterers in the near-field region, even though the physical array size is small.

\section{Cross-field MIMO Channel Model Generation} \label{sec:simulation}

In this section, detailed generation procedures of the cross-field MIMO channel model are described, and simulation results are discussed.

\subsection{Generation Procedure} \label{sec:procedure}
\par
\begin{figure*}
    \centering
    \includegraphics[width=\linewidth]{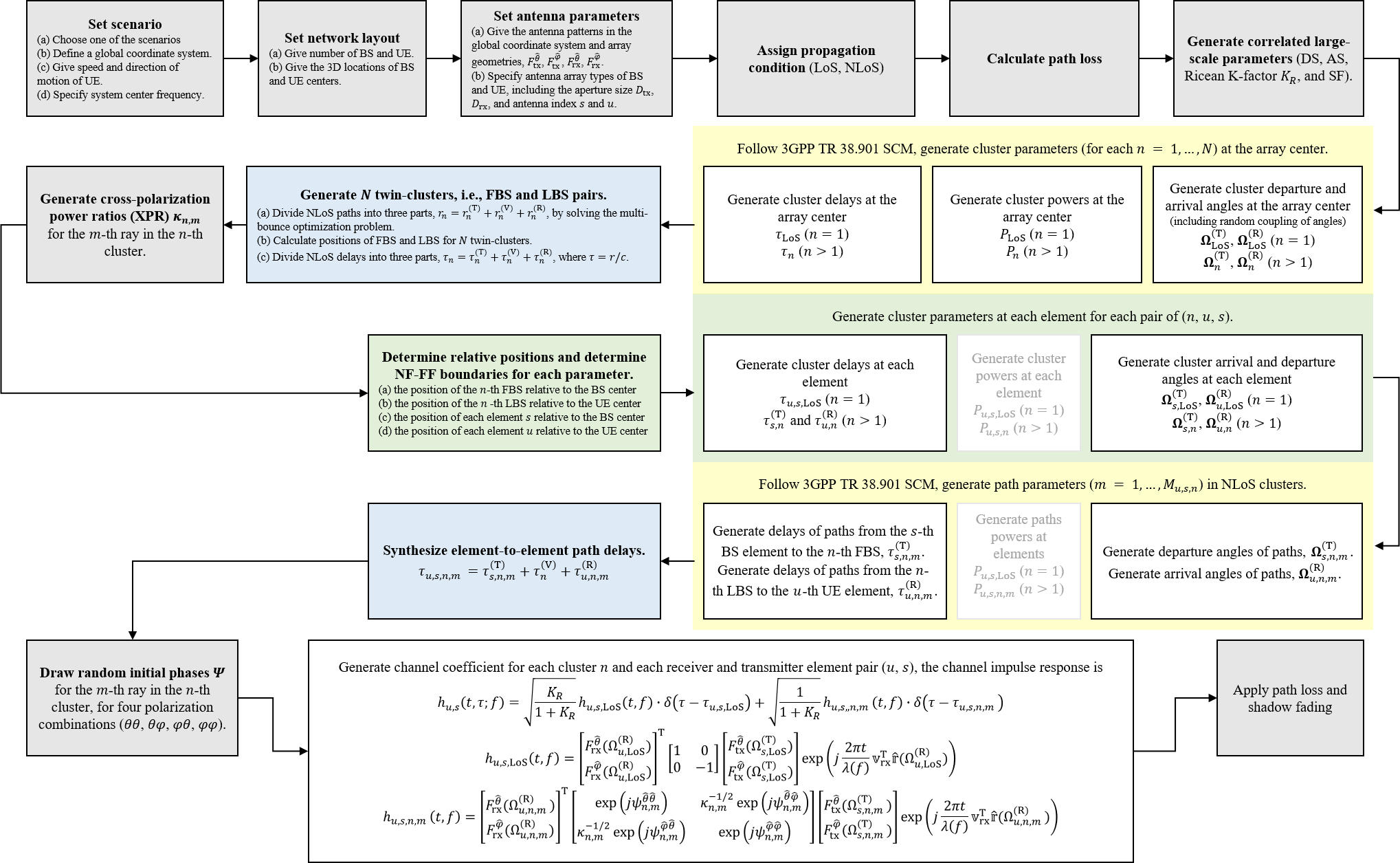}
    \caption{Schematic diagram of cross-field MIMO channel model. Blue blocks are new based on the twin-scatterer framework. Yellow and green blocks are split from steps in the 3GPP TR~38.901 channel model. The green part distinguishes NF and FF MPC parameters. Grey blocks remain the same with the 3GPP TR~38.901 channel model.}
    \label{fig:procedure}
\end{figure*}

 As illustrated in Figure~\ref{fig:procedure}, the procedure is based on the model in 3GPP TR~38.901, while modified steps are described as follows.

Step 7: Follow 3GPP TR 38.901 SCM, generate cluster parameters ($n = 1, ..., N$), while includes (a) cluster delays, $\tau_{1,{\rm LoS}}$ and $\tau_{n}$, (b) cluster powers, $P_{1,{\rm LoS}}$ and $P_{n}$, and (c) cluster departure and arrival angles (including random coupling of angles), $\Omega_{\rm 1,LoS}^{\rm(T)}$, $\Omega_{\rm 1,LoS}^{\rm(T)}$, $\Omega_{n}^{\rm(R)}$, $\Omega_{n}^{\rm(R)}$.

Step 8: Generate $N$ twin scatterers, i.e., first-bounce-scatterer (FBS) and last-bounce-scatterer (LBS) pairs, while includes
(a) divide NLoS paths into three parts, $r_n=r_{n}^{\rm(T)}+r_{n}^{\rm(V)}+r_{n}^{\rm(R)}$, by solving the multi-bounce optimization problem.
(b) calculate positions of FBS and LBS for $N$ twin scatterers,
and (c) divide NLoS delays into three parts, $\tau_n=\tau_{n}^{\rm(T)}+\tau_{n}^{\rm(V)}+\tau_{n}^{\rm(R)}$, where $\tau=r/c$.

Step 10: Determine relative positions in the local coordinate system where the UPA lies in the $xy$-plane, including
(a) the position of the $n$$^\mathrm{th}$ FBS relative to the BS center is denoted as $\textbf{r}_{n}^{\rm(T)}$, which is determined by length $r_{n}^{\rm(T)}=c\cdot\tau_{n}^{\rm(T)}$ and angles $\Omega_{n}^{\rm(T)}$. The corresponding unit vector is denoted as $\hat{\textbf{r}}_{n}^{\rm(T)}$, and $\textbf{r}_{n}^{\rm(T)}=r_{n}^{\rm(T)}\cdot\hat{\textbf{r}}_{n}^{\rm(T)}$,
(b) the position of the $n$$^\mathrm{th}$ LBS relative to the UE center is denoted as $\textbf{r}_{n}^{\rm(R)}$, which is determined by length $r_{n}^{\rm(R)}=c\cdot\tau_{n}^{\rm(R)}$ and angles $\Omega_{n}^{\rm(R)}$. The corresponding unit vector is denoted as $\hat{\textbf{r}}_{n}^{\rm(R)}$, and $\textbf{r}_{n}^{\rm(R)}=r_{n}^{\rm(R)}\cdot\hat{\textbf{r}}_{n}^{\rm(R)}$,
(c) the position of each element $s$ relative to the BS center, $\textbf{r}_{s}^{\rm(T)}$, is determined by the array structure,
and (d) the position of each element $u$ relative to the UE center, $\textbf{r}_{u}^{\rm(R)}$, is determined by the array structure.

Step 11: Generate cluster delays at each element, $\tau_{u,s,{\rm LoS}}$, $\tau_{s,n}^{\rm(T)}$ and $\tau_{u,n}^{\rm(R)}$.
For the distance between the $n$$^\mathrm{th}$ FBS and the $s$$^\mathrm{th}$ BS element, $r_{s,n}^{\rm(T)}$, and the distance between the $n$$^\mathrm{th}$ LBS and the $u$$^\mathrm{th}$ UE element, $r_{u,n}^{\rm(R)}$, conduct the following steps respectively.
(Take $r_{s,n}^{\rm(T)}$ as example)
If $r_{n}^{\rm(T)}>\frac{2\cdot{D_{\rm tx}}^2}{\lambda}$,
\begin{equation}
    r_{s,n}^{\rm(T)} \approx r_{n}^{\rm(T)} - \hat{\textbf{r}}_{n}^{\rm(T)} \cdot \textbf{r}_{s}^{\rm(T)}
\end{equation}
otherwise, if $r_{n}^{\rm(T)}>0.62\cdot\sqrt{\frac{{D_{\rm tx}}^3}{\lambda}}$,
\begin{equation}
    r_{s,n}^{\rm(T)} \approx r_{n}^{\rm(T)} - \hat{\textbf{r}}_{n}^{\rm(T)} \cdot \textbf{r}_{s}^{\rm(T)} + \frac{{\left\|\textbf{r}_{s}^{\rm(T)}\right\|}^2-{\left(\hat{\textbf{r}}_{n}^{\rm(T)} \cdot \textbf{r}_{s}^{\rm(T)}\right)}^2}{2\cdot r_{n}^{\rm(T)}}
\end{equation}
otherwise, accurately calculate $r_{s,n}^{\rm(T)}$.
Finally, $\tau_{s,n}^{\rm(T)}=r_{s,n}^{\rm(T)}/c$.

Step 12: Generate cluster arrival and departure angles at each element $\Omega_{s,{\rm LoS}}^{\rm(T)}$, $\Omega_{u,{\rm LoS}}^{\rm(R)}$, $\Omega_{s,n}^{\rm(T)}$, and $\Omega_{u,n}^{\rm(R)}$.
For the elevation angle of departure from the $s$$^\mathrm{th}$ BS element to the $n$$^\mathrm{th}$ FBS $\theta_{s,n}^{\rm(T)}$ and the elevation angle of arrival from the $n$$^\mathrm{th}$ LBS to the $u$$^\mathrm{th}$ UE element $\theta_{u,n}^{\rm(R)}$. (Take $\theta_{s,n}^{\rm(T)}$ as example)
    If 
    \begin{equation*}
        r_{n}^{\rm(T)} > \frac{D_{\rm tx}}{2} \cdot \left[\sin\theta_{n}^{\rm(T)}-\tan(\theta_{n}^{\rm(T)}-\frac{{\rm HPBW}_{\rm tx}^{\rm(V)}}{2})\cdot\cos\theta_{n}^{\rm(T)}\right]
    \end{equation*}
    then
    \begin{equation}
        \theta_{s,n}^{\rm(T)}=\theta_{n}^{\rm(T)}
    \end{equation}
    otherwise, accurately calculate
    \begin{equation}
        \theta_{s,n}^{\rm(T)}=\arctan\frac{\left(\left[\textbf{r}_{n}^{\rm(T)}-\textbf{r}_{s}^{\rm(T)}\right]_x^2+\left[\textbf{r}_{n}^{\rm(T)}-\textbf{r}_{s}^{\rm(T)}\right]_y^2\right)^{1/2}}{\left[\textbf{r}_{n}^{\rm(T)}-\textbf{r}_{s}^{\rm(T)}\right]_z^2}
    \end{equation}
    where $\left[\cdot\right]_x$, $\left[\cdot\right]_y$, $\left[\cdot\right]_z$ represent the coordinate of the vector in the Cartesian LCS.
For the azimuth angle of departure from the $s$$^\mathrm{th}$ BS element to the $n$$^\mathrm{th}$ FBS $\varphi_{s,n}^{\rm(T)}$ and the azimuth angle of arrival from the $n$$^\mathrm{th}$ LBS to the $u$$^\mathrm{th}$ UE element $\varphi_{u,n}^{\rm(R)}$. (Take $\varphi_{s,n}^{\rm(T)}$ as example)
    If 
    \begin{equation*}
    \begin{aligned}
        \max_{(s)\ \rm at\ vertice} \left|\varphi_{n}^{\rm(T)}-\arctan\frac{\left[\textbf{r}_{n}^{\rm(T)}-\textbf{r}_{s}^{\rm(T)}\right]_y}{\left[\textbf{r}_{n}^{\rm(T)}-\textbf{r}_{s}^{\rm(T)}\right]_x}\right| < \frac{{\rm HPBW}_{\rm tx}^{\rm(H)}}{2}
    \end{aligned}
    \end{equation*}
    then
    \begin{equation}
        \varphi_{s,n}^{\rm(T)}=\varphi_{n}^{\rm(T)}
    \end{equation}
    otherwise, accurately calculate
    \begin{equation}
        \varphi_{s,n}^{\rm(T)}=\arctan\frac{\left[\textbf{r}_{n}^{\rm(T)}-\textbf{r}_{s}^{\rm(T)}\right]_y}{\left[\textbf{r}_{n}^{\rm(T)}-\textbf{r}_{s}^{\rm(T)}\right]_x}
    \end{equation}

Step 13: Follow 3GPP TR 38.901 SCM, generate path parameters ($m = 1, ..., M_{u,s,n}$) in NLoS clusters at each ($n$,$u$) pair and ($n$,$s$) pair, including
(a) delays of paths ($m = 1, ..., M_{u,s,n}$) from the $s$$^\mathrm{th}$ BS element to the $n$$^\mathrm{th}$ FBS, $\tau_{s,n,m}^{\rm(T)}$, (b) delays of paths ($m = 1, ..., M_{u,s,n}$) from the $n$$^\mathrm{th}$ LBS to the $u$$^\mathrm{th}$ UE element, $\tau_{u,n,m}^{\rm(R)}$, (c) departure angles of paths, $\Omega_{s,n,m}^{\rm(T)}$, and (d) arrival angles of paths, $\Omega_{u,n,m}^{\rm(R)}$.

Step 14: Synthesize element-to-element path delays by $\tau_{u,s,n,m}=\tau_{s,n,m}^{\rm(T)}+\tau_{n}^{\rm(V)}+\tau_{u,n,m}^{\rm(R)}$.



\subsection{Simulation Results} \label{sec:result}

\begin{figure}[t]
    \centering
    \begin{subfigure}[Location of Tx, Rx, and twin scatterers.]{
    \includegraphics[width=0.9\linewidth]{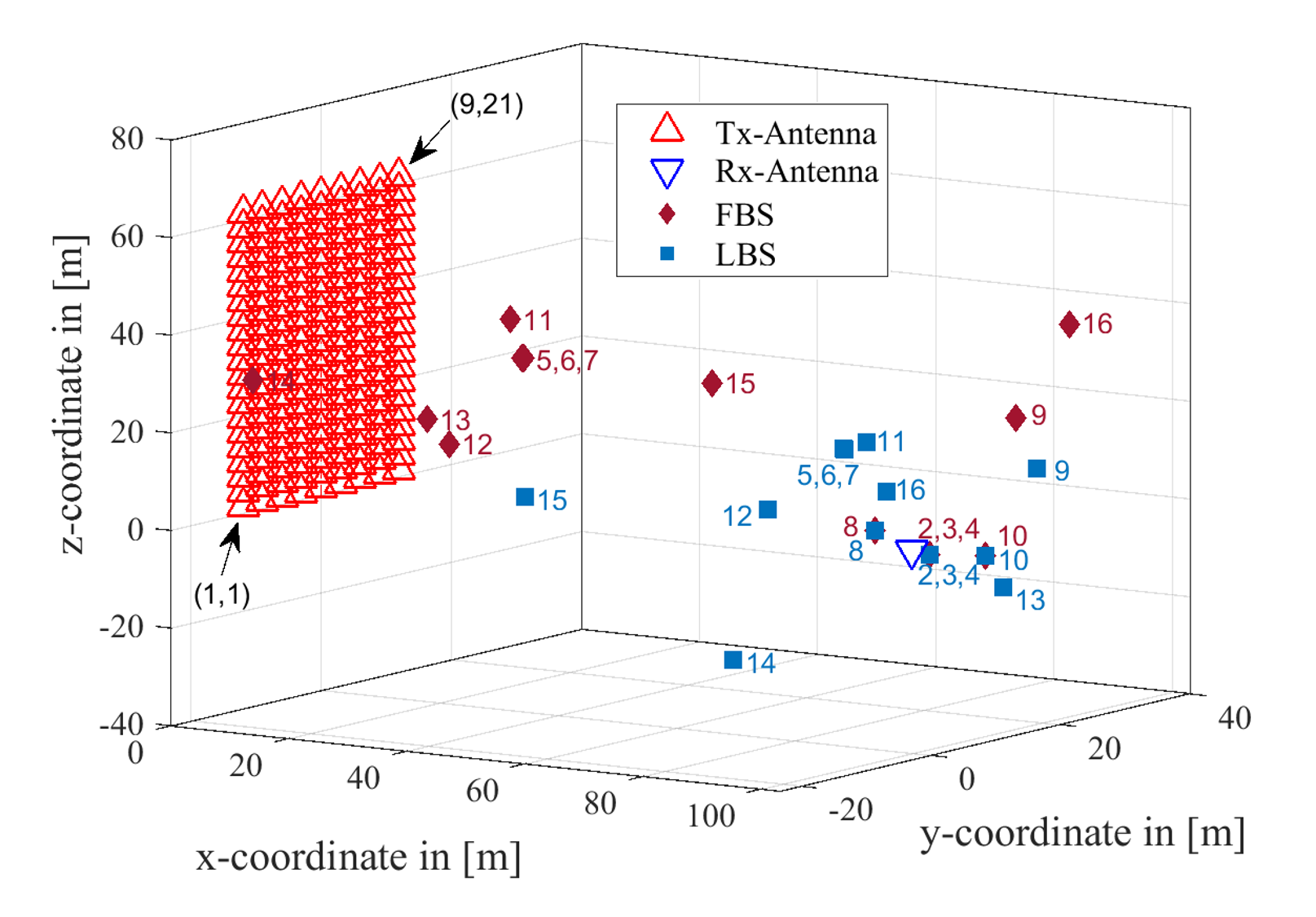}}
    \end{subfigure}
    \\
    \begin{subfigure}[PDAP.]{
    \includegraphics[width=0.7\linewidth]{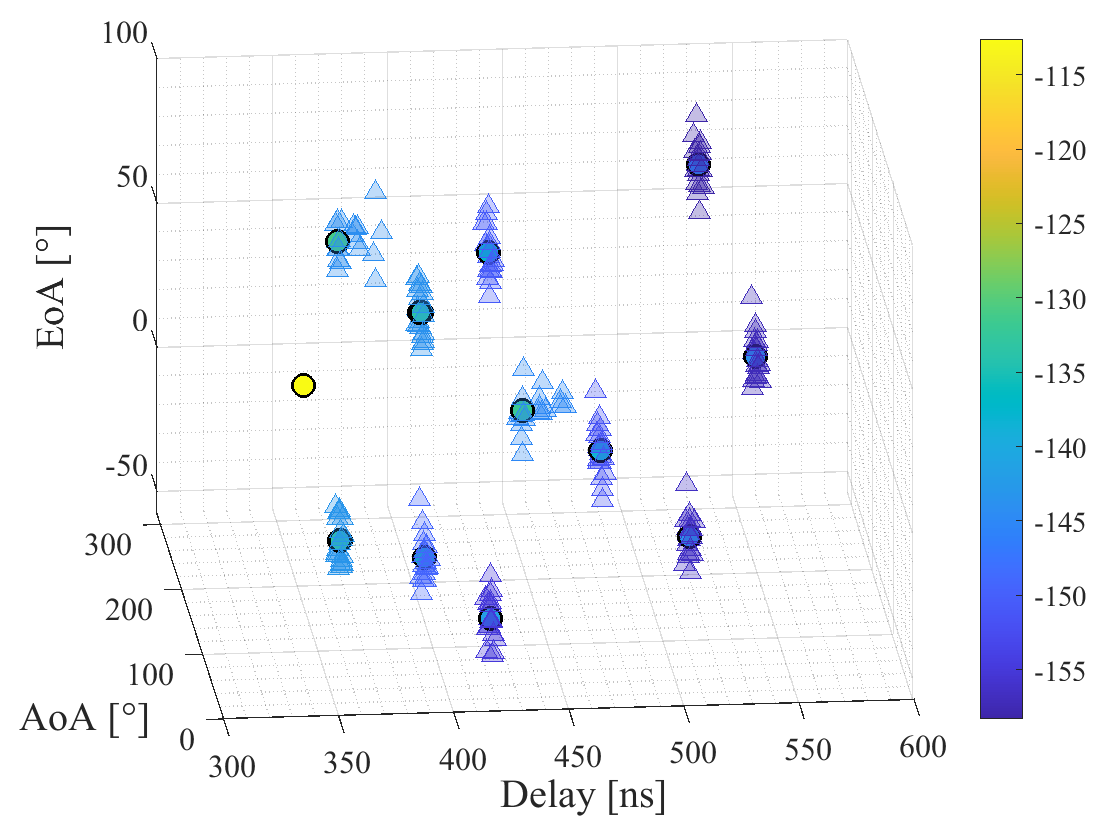}}
    \end{subfigure}
    \caption{An example of the simulation result at 140~GHz. 9$\times$21 UPA is located in the $yz$-plane, and $d_{\rm 2D}=100$~m. (a) Location of Tx, Rx, and twin scatterers. Since the first cluster is the LoS ray, numbers 2-16 are used to label 15 pairs of twin scatterers. A pair of overlapped LBS and FBS represents the scatterer location of a one-bounce path. (b) PDAP result at the (5,1)$^\mathrm{th}$ antenna element. Circles represent the major MPC in clusters, while triangles denote other MPCs in the cluster. The color bar shows the path gain in dB.}
    \label{fig:simulation_result_example}
\end{figure}

\begin{figure}
    \centering
    \begin{subfigure}[FF probability in terms of EoD.]{
    \includegraphics[width=\linewidth]{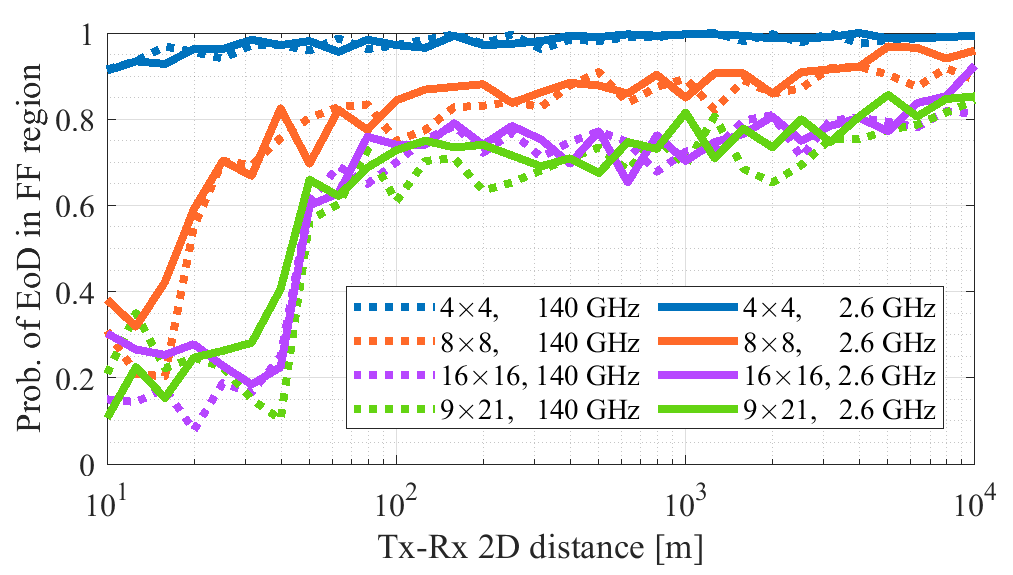}}
    \end{subfigure}
    \\
    \begin{subfigure}[FF probability in terms of delay.]{
    \includegraphics[width=\linewidth]{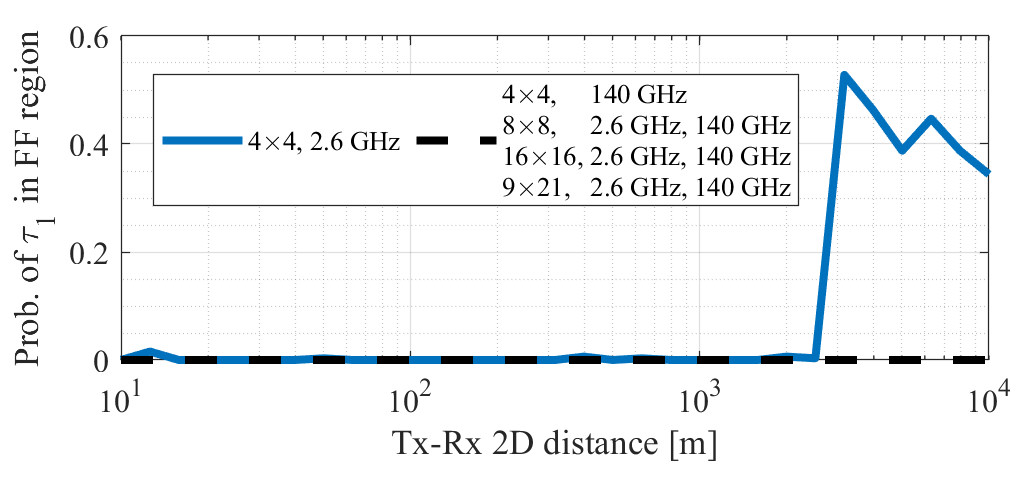}}
    \end{subfigure}
    \caption{The probability that an FBS is in the FF region in terms of (a) EoD or (b) delay, under different 2D Tx-Rx distances at 2.6~GHz and 140~GHz.}
    \label{fig:summary_spacing3m}
\end{figure}

\par We first measures the statistical probabilities for the scatter to be in the near-field of the antenna array, in terms of MPC parameters, namely the delay and angles. Following the procedure described in Section~\ref{sec:procedure}, we simulate the MISO channel in a UMa scenario at 2.6~GHz and 140~GHz. The Rx's height is 2.5~m. Four types of antenna array are equipped at Tx respectively, namely $4\times4$, $8\times8$, $16\times16$, and $9\times21$ UPA. The antenna element spacing is 3~m, much larger than the wavelength, and thus the largest two ELAAs span an area of 45~m $\times$ 45~m, and 24~m $\times$ 60~m.
The center of the Tx array is 25~m high for $4\times4$, $8\times8$, and $16\times16$ UPA. The $9\times21$ UPA represents the scenario where antennas are hidden in construction elements~\cite{bjornson2019massive}. In this case, the height of the Tx array center is 31.5~m to ensure that every antenna element is higher than Rx. We simulate each antenna element as a custom antenna with 3-dB beamwidth of $120^\circ$ both vertically and horizontally.

\par Fig.~\ref{fig:simulation_result_example} shows one of the simulation results. In each trial, 15 pair of twin scatterers are generated, and the simulation with the same setup is repeated by 20 times. We calculate the statistical probability for the FBS located in the FF region, with respect to the path length from Tx antenna to FBS (or $\tau_1$), the azimuth angle of departure (AoD), and the elevation angle of departure (EoD), respectively. The 2D Tx-Rx distance, $d_{\rm 2D}$ varies from 10~m to 10~km.

\par The results are summarized as follows.
First, the FBS is always in the FF region in terms of AoD, and the value of AoD can be regarded as the same across the ELAA at either 2.6~GHz or 140~GHz.
\par Second, the probability that an FBS is in the FF region in terms of EoD becomes higher as Rx gets away from Tx. It is also verified that the NF-FF boundary in terms of angles is unrelated with the carrier frequency. More specifically, as shown in Fig.~\ref{fig:summary_spacing3m}(a), at a certain range of $d_{\rm 2D}$, the FF probability rises rapidly and then saturates. Besides, the larger the antenna aperture size, the lower the likelihood that an FBS is in the FF region where the EoD from the Tx to the FBS should be calculated element-by-element. For instance, for ELAA than spans 45~m $\times$ 45~m ($16\times16$ UPA) or 24~m $\times$ 60~m ($9\times21$ UPA), when the 2D Tx-Rx distance $d_{\rm 2D}$ is 30~m, 70\%-80\% of the FBS are in the NF region.
\par Third, as shown in Fig.~\ref{fig:summary_spacing3m}(b), in all cases at 140~GHz, FBS is in the NF region of the Tx antenna array in terms of delay or path length. By contrast, at 2.6~GHz, the large wavelength narrows down the NF region. Even though, for the antenna aperture larger than 21~m $\times$ 21~m ($8\times8$ UPA), all simulated FBS are located in the near-field region. In this case, the path length (or propagation time $\tau_1$) between the Tx antenna and the FBS should be calculated with a more accurate approximation. When the antenna aperture is reduced to 9~m $\times$ 9~m ($4\times4$ UPA), and the 2D distance between Tx and Rx is larger than 3~km, there is a 40\%-50\% likelihood that an FBS falls in the FF region in terms of the path length (or propagation time $\tau_1$).


\begin{figure}
    \centering
    \begin{subfigure}[Tx is 4$\times$4, at 2.6~GHz.]{
    \includegraphics[width=0.47\linewidth]{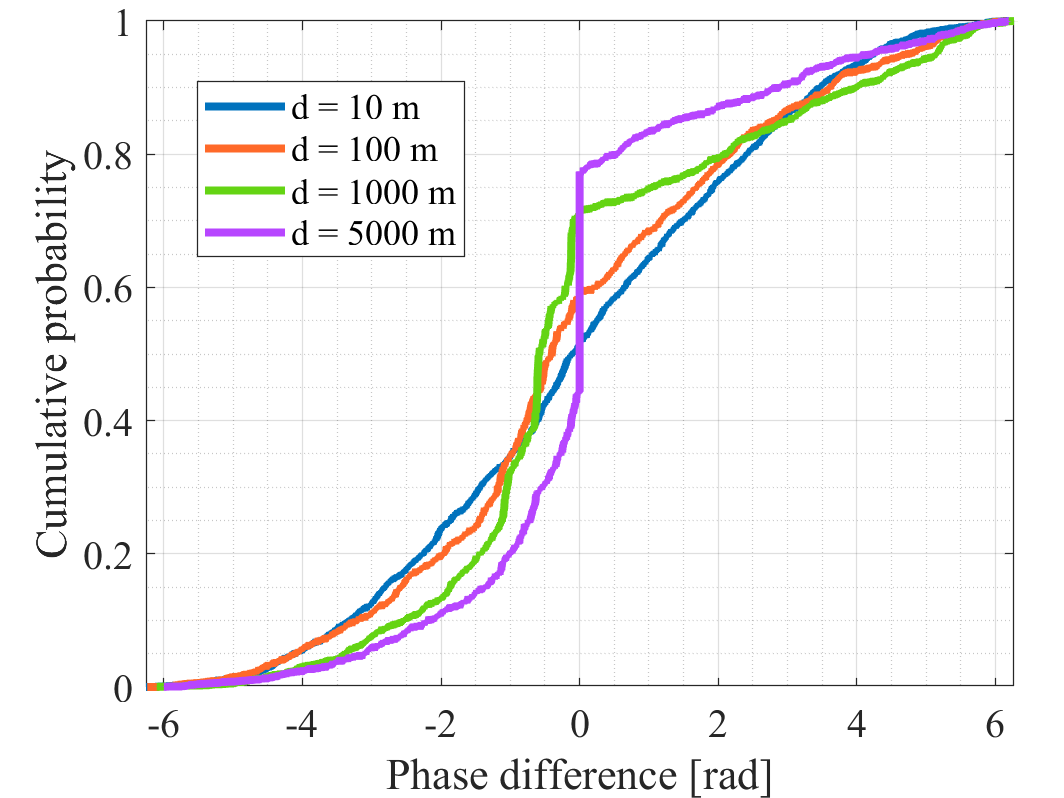}}
    \end{subfigure}
    \begin{subfigure}[Tx is 16$\times$16, at 2.6~GHz.]{
    \includegraphics[width=0.47\linewidth]{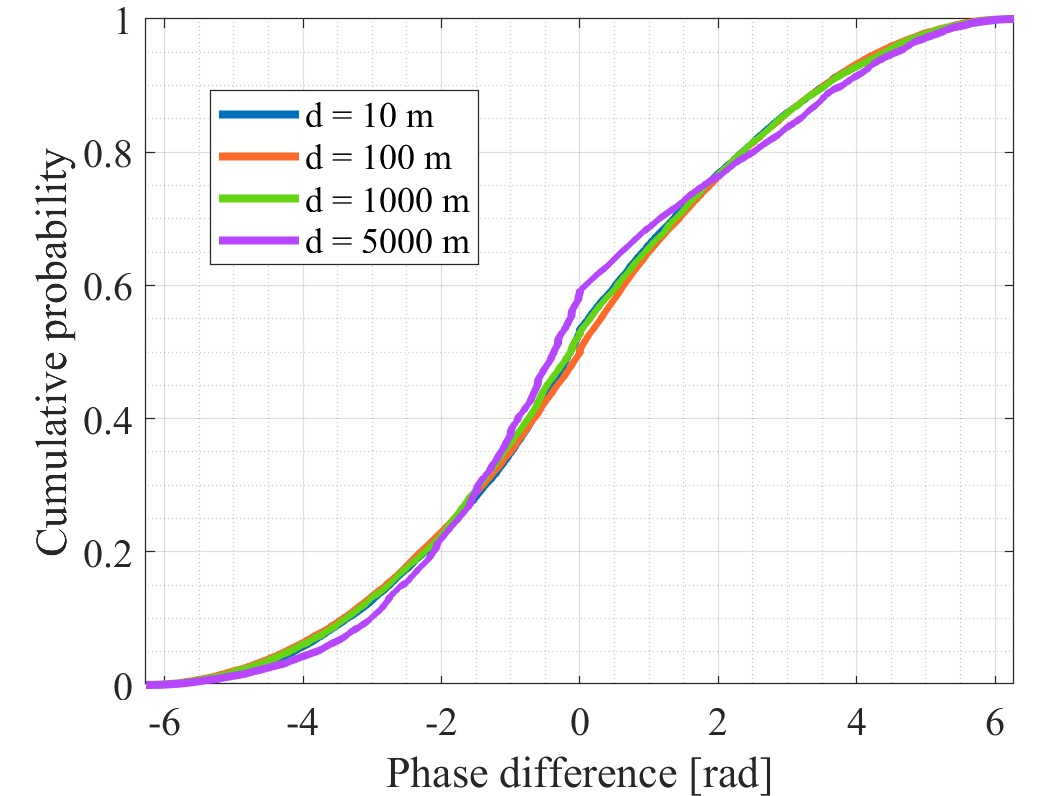}}
    \end{subfigure}
    \\
    \begin{subfigure}[Tx is 4$\times$4, at 140~GHz.]{
    \includegraphics[width=0.47\linewidth]{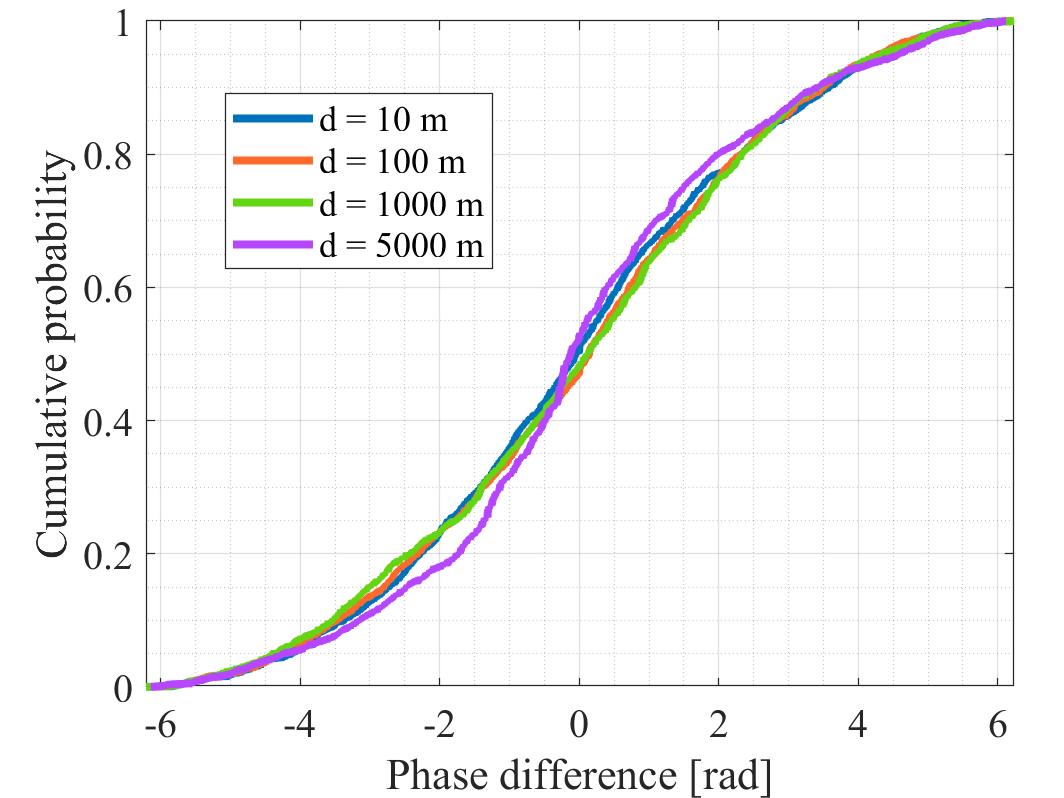}}
    \end{subfigure}
    \begin{subfigure}[Tx is 16$\times$16, at 140~GHz.]{
    \includegraphics[width=0.47\linewidth]{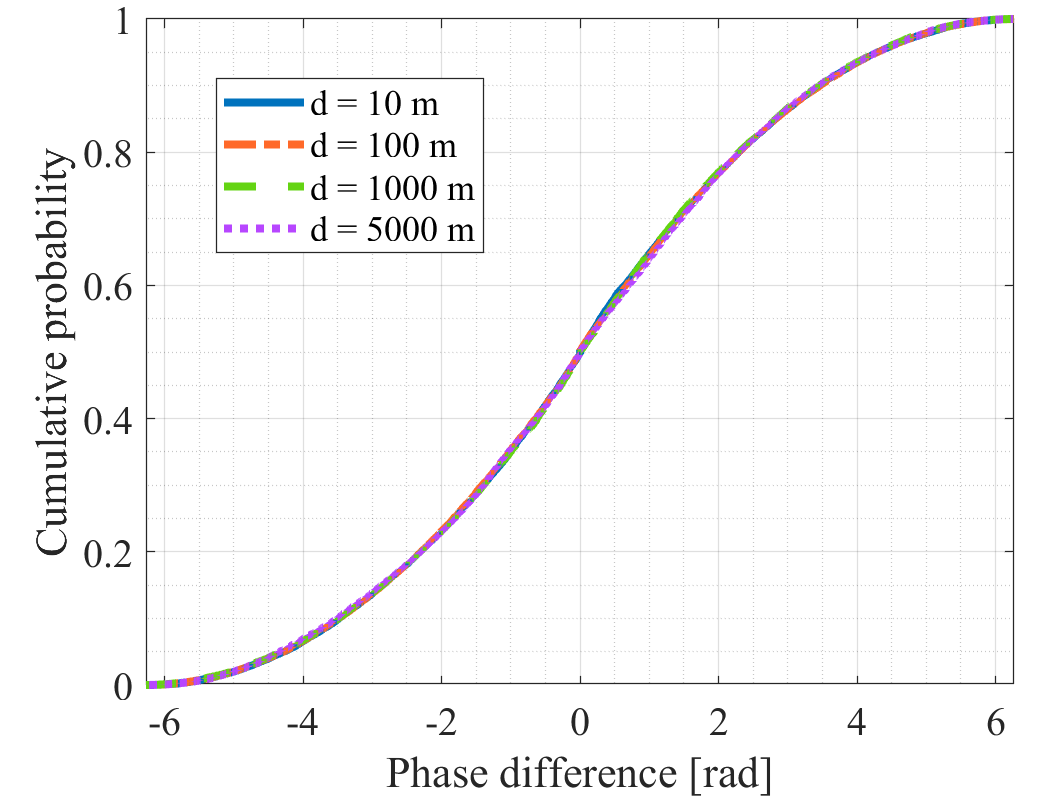}}
    \end{subfigure}
    \caption{Phase difference between far-field characterization and cross-field characterization in the UMa scenario ($h_{\rm Tx}=25$~m, $h_{\rm Rx}=1.5$~m) at different frequencies and 2D distance $d$ between Tx and Rx.}
    \label{fig:phase_diff_sim}
\end{figure}

Then, we compare the proposed cross-field modeling with the far-field modeling of channels in ELAA systems. Fig.~\ref{fig:phase_diff_sim} shows the difference of phase for all paths and element-to-element links between CF and FF characterization. First, as the 2D distance between Tx and Rx increases, the phase difference between cross-field and pure far-field model is attenuated since there is a higher probability for scatterers to be located in the far-field. Second, as compared in Fig.~\ref{fig:phase_diff_sim}(a)(b) and (c)(d), the larger the antenna array, the more the deviation between CF and FF modeling. Third, as compared in Fig.~\ref{fig:phase_diff_sim}(a)(c) and (b)(d), the higher the frequency, the more the deviation between CF and FF modeling. Overall, as the cross-field phenomena become significant for ELAA systems, the distribution of phase difference tends to the normal distribution between $-2\pi$ and $2\pi$.

\begin{figure}
    \centering
    \begin{subfigure}[Tx is 4$\times$4, at 2.6~GHz.]{
    \includegraphics[width=0.47\linewidth]{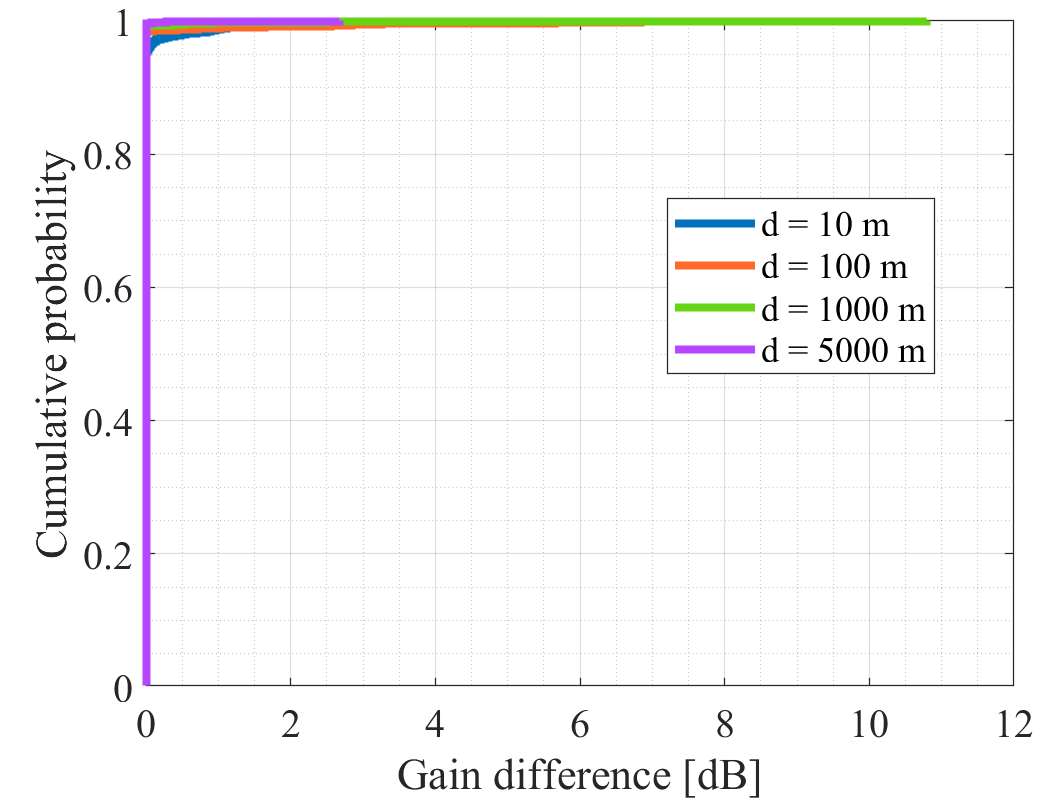}}
    \end{subfigure}
    \begin{subfigure}[Tx is 16$\times$16, at 2.6~GHz.]{
    \includegraphics[width=0.47\linewidth]{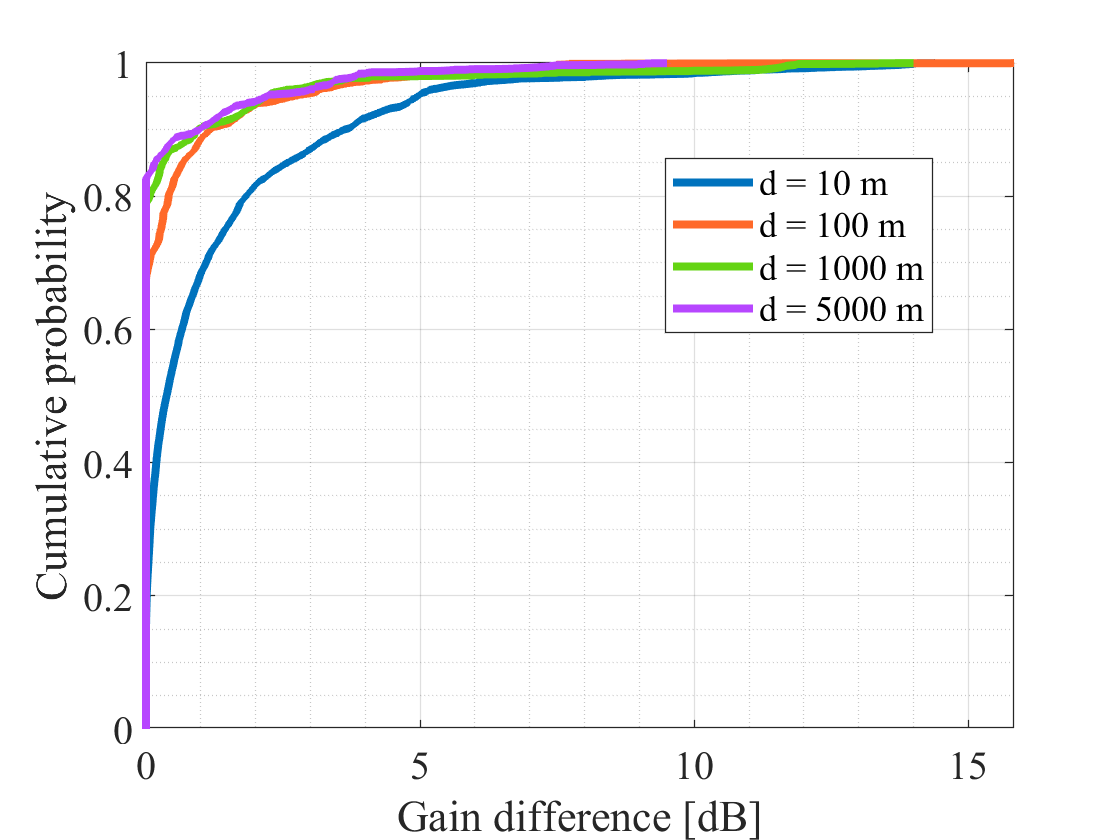}}
    \end{subfigure}
    \\
    \begin{subfigure}[Tx is 4$\times$4, at 140~GHz.]{
    \includegraphics[width=0.47\linewidth]{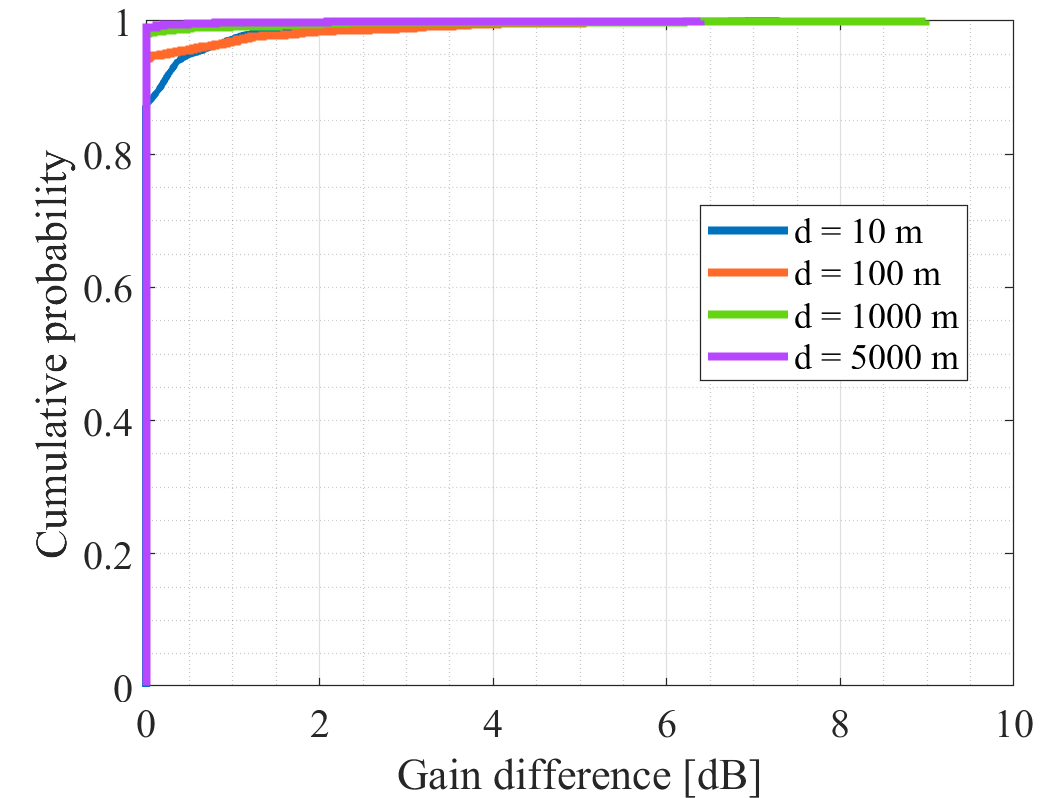}}
    \end{subfigure}
    \begin{subfigure}[Tx is 16$\times$16, at 140~GHz.]{
    \includegraphics[width=0.47\linewidth]{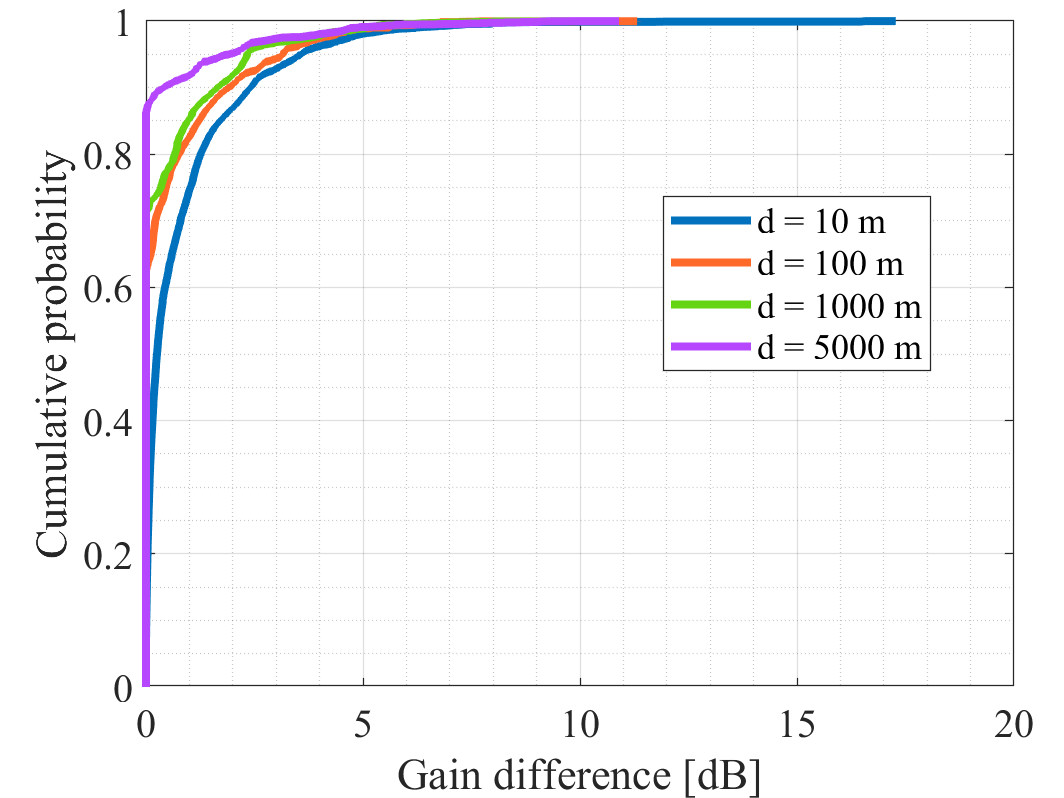}}
    \end{subfigure}
    \caption{Gain difference between far-field characterization and cross-field characterization in the UMa scenario ($h_{\rm Tx}=25$~m, $h_{\rm Rx}=1.5$~m) at different frequencies and 2D distance $d$ between Tx and Rx.}
    \label{fig:gain_diff_sim}
\end{figure}

For the difference of gain for all paths and element-to-element links between CF and FF, as illustrated in Fig.~\ref{fig:gain_diff_sim}. The size of antenna array is more significant then the frequency band. Still, the farther then distance, the less the difference between CF and FF characterization.
\section{Conclusion} \label{sec:conclusion}

In this paper, we propose a cross near- and far-field MIMO channel model for ELAA systems.
This starts with the introduction of the channel model framework based on twin-scatterer, i.e., the EM wave propagation is divided into three parts, separated by its first and last bounce.
Then we provide a comprehensive analysis and closed-form expressions of cross-field boundaries and MPC parameters. By comparing the distance between the scatterer and the antenna array with the NF-FF boundary, far-field and near-field MPC parameters are generated separately.
The analysis is verified by a cross-field channel measurement carried out in a small indoor scenario at 300~GHz, which also demonstrates the necessity of element-level modeling of MIMO channels when we have scatterers located in the near-field region.
Furthermore, detailed channel generation procedures are provided, and simulations are conducted in a UMa scenario with $4\times4$, $8\times8$, $16\times16$, and $9\times21$ UPAs at Tx at 2.6~GHz and 140~GHz. It turns out that element-level characterization of path length (or propagation time) in the near-field is indispensable for ELAA systems. Even for small antenna aperture like 9~m $\times$ 9~m and 2D Tx-Rx distance larger than 3~km, there is also a 50\%-60\% likelihood that a scatterer falls in the near-field region in terms of the path length. By contrast, at least 70\%-80\% of elevation angle is in the near-field for ELAA systems that span as large as 45~m $\times$ 45~m or 24~m $\times$ 60~m when the 2D Tx-Rx distance is less than 30~m.
Moreover, channel coefficients are compared between the proposed cross-field model and the original far-field model. While the difference in gain is limited to several decibels, the phase could deviate by at most a whole period of $2\pi$, which motivates the careful characterization of MPC parameters in cross-field modeling.



\bibliographystyle{IEEEtran}
\bibliography{bibliography}

\end{document}